\documentclass[prl,aps,amssymb,amsmath]{article}
\usepackage{amssymb,amsmath}
\newtheorem{twr}{Theorem}
\newtheorem{lem}{Lemma}

\begin{document}
\date{June 11, 2003}
\title{{\bf A Generalization of the Bargmann's Theory of Ray Representations}}
\author{Jaros{\l}aw Wawrzycki \footnote{Electronic address: 
Jaroslaw.Wawrzycki@ifj.edu.pl or  
jwaw@th.if.uj.edu.pl}
\\Institute of Nuclear Physics, ul. Radzikowskiego 152, 
\\31-342 Krak\'ow, Poland}
\maketitle
\newcommand{\ud}{\mathrm{d}}
\begin{abstract}
The paper contains a complete theory of factors for ray representations
acting in a Hilbert bundle, which is a generalization
of the known Bargmann's theory. With the help of it we have reformulated 
the standard quantum theory such that the gauge freedom emerges naturally
from the very nature of quantum laws. The theory is of primary importance
in the investigations of covariance (in contradistinction to symmetry) 
of a quantum theory which possesses a nontrivial gauge freedom.
In that case the group in question is not any symmetry group but it is a 
covariance group only -- that case 
which has not been deeply investigated.
It is shown in the paper that the factor of its representation depends on 
space and time when the system in question possesses a gauge freedom. In the 
nonrelativistic theories the factor depends on the time only.
In the relativistic theory the Hilbert bundle is over the spacetime and in the 
nonrelativistic one it is over the time.
 
We explain two applications of this generalization: 
in the theory of a quantum particle in the gravitational field in the 
nonrelativistic limit and in the quantum electrodynamics.
\end{abstract}
 
\vspace{1cm}

\section{Introduction}

In the standard Quantum Mechanics (QM) and the Quantum Field Theory (QFT) 
the spacetime coordinates are pretty classical variables. Therefore
the question about the general covariance of QM and QFT
emerges naturally just like in the classical theory:
\begin{enumerate} 

\item[]
        what is the effect of a changing of the spacetime coordinates
        in QM and QFT when the changing does not form any symmetry
        transformation?

\end{enumerate}
It is a commonly accepted believe that there are no substantial 
difficulties if we refer the question to the wave equation. We simply
treat the wave equation, and do not say why, in such a manner as if it 
was a classical equation. The only problem arising is to find the 
transformation rule $\psi \to T_{r}\psi$ for the wave function $\psi$.
This procedure, which on the other hand can be seriously objected, does
not solve the above stated problem. The heart of the problem as well
as of QM and QFT lies in the Hilbert space of states and just in 
finding the representation $T_{r}$ of the covariance group in question.
The trouble gets its source in the fact that the covariance transformation
changes the form of the wave equation such that $\psi$ and $T_{r}\psi$
do not belong to the same Hilbert space, which means that $T_{r}$ does not
act in the ordinary Hilbert space. This is not compatible with the 
paradigm worked out in dealing with symmetry groups. 

We show that covariance group acts in a Hilbert bundle
$\mathcal{R}\triangle \mathcal{H}$ over the time in the nonrelativistic 
theory and in a Hilbert bundle $\mathcal{M}\triangle \mathcal{H}$ over
the spacetime $\mathcal{M}$ in the relativistic case. The waves are
the appropriate cross sections of the bundle in question. The exponent
$\xi(r,s,p)$ in the formula
\begin{displaymath}
T_{r}T_{s} = e^{i\xi(r,s,p)}T_{rs},
\end{displaymath}  
depends on the point $p$ of the base of the bundle in question: 
that is, $\xi$ depends on the time $t$ in the nonrelativistic 
theory and on spacetime point $p$ in the relativistic theory if 
there exists a nontrivial gauge freedom. 

Moreover, we argue that the bundle $\mathcal{M}\triangle \mathcal{H}$
is more appropriate for treating the covariance as well as the symmetry
groups then the Hilbert space itself. Namely, we show that from the 
more general assumption that the representation $T_{r}$ of the Galilean
group acts in $\mathcal{R}\triangle \mathcal{H}$ and has an exponent
$\xi(r,s,t)$ depending on the time $t$ we reconstruct the nonrelativistic
Quantum Mechanics. Even more, in the less trivial case of the theory
with nontrivial time-dependent gauge describing the spin less quantum 
particle in the Newtonian gravity we are able to infer the wave equation
and prove the equality of the inertial and gravitational masses.  
  
In doing it we apply extensively the classification theory for exponents
$\xi(r,s,t)$ of $T_{r}$ acting in $\mathcal{R}\triangle \mathcal{H}$ 
and depending on the time. 

The main task of this paper is to construct the general classification
theory of spacetime dependent exponents $\xi(r,s,p)$ of representations
acting in $\mathcal{M}\triangle \mathcal{H}$. On the other hand 
the presented theory can be viewed as a generalization of the 
Bargmann's \cite{Bar} classification theory of exponents $\xi(r,s)$
of representations acting in ordinary Hilbert spaces,
which are independent of $p \in \mathcal{M}$.  

In the presented theory which is slightly more general then the standard one
the gauge freedom emerges from the very nature of the fundamental laws of
Quantum Mechanics. By this it opens a new perspective in solving the
troubles in QFT caused by the gauge freedom. 
 
In section {\bf 2} we present the physical motivation in detail. In section 
{\bf 3} we present
the generalization of the ordinary state vector ray 
and operator ray introduced by H. Weyl. In sections {\bf 3} and {\bf 4} we 
present the continuity assumption from which the strong continuity of the 
exponent $\xi$ follows and generalize the ordinary notion of the  
exponent $\xi$ of a ray representation. In section {\bf 5} we analyze the 
local exponents of Lie groups. In section {\bf 6} we introduce algebras which 
are the important tools for the classification theory of local exponents 
presented in the section {\bf 7}. In section {\bf 8} we investigate the 
globally defined exponents and classify them in some special cases.
In the section {\bf 9} we present examples. The first example is the  
Galilean group. We analyze the group from the point of view of the generalized 
theory. As the second example the exponents of the Milne group, 
the covariance group relevant in the theory of nonrelativistic 
particle in the gravitational field, are analyzed.

\vspace{1ex}

The proof of differentiability of the (generalized) exponent and the first
three Lemmas goes in an analogous way as those presented in the Bargmann's 
work \cite{Bar}.
 However, it is not trivial that they are also true in this 
generalized situation.
 We present the proof of them explicitly for the 
reader's convenience. The rest of our
 reasoning is not a simple analogue of 
\cite{Bar} and proceeds another way.

\section{Setting for the Motivation}\label{motivation}

In this subsection we carry out the general analysis of the representation 
$T_{r}$ of a covariance
group and compare it with the representation of a 
symmetry group. We describe also the correspondence between the space of wave 
functions $\psi(\vec{x},t)$ and the Hilbert space. We carry out the analysis 
in the nonrelativistic case, but
it can be derived as well in the relativistic 
quantum field theory. 

Before we give the general description, it will be instructive to investigate 
the problem for the free particle in the flat Galilean spacetime. The set of 
solutions $\psi$ of the Schr\"odinger equation which are admissible in Quantum 
Mechanics is 
precisely given by 
\begin{displaymath}
\psi(\vec x,t)=(2\pi)^{-3/2} \int \varphi(\vec k)e^{-i\frac{t}{2m}\vec{k}
\centerdot \vec k+i\vec k\centerdot\vec x}
\, {\ud}^{3}{k},
\end{displaymath}
where $p=\hslash k$ is the linear momentum and $\varphi(\vec{k})$ is any square 
integrable function. The functions $\varphi$ (wave functions in the "Heisenberg 
picture") form a Hilbert space $\mathcal{H}$ with the inner product
\begin{displaymath}
(\varphi_{1}, \varphi_{2})=\int \varphi_{1}^{*}(\vec{k})\varphi_{2}(\vec{k}) \, 
{\ud}^{3}{k}.
\end{displaymath}
The correspondence between $\psi$ and $\varphi$ is one-to-one. 

But in general the construction fails if the Schr\"odinger equation possesses 
a nontrivial gauge freedom.
We explain it. For example the above construction 
fails for the nonrelativistic
quantum particle in the curved 
Newton-Cartan 
spacetime. Beside this, in this spacetime we do not have plane wave, see 
\cite{Waw}. 
So, there does not exist any natural counterpart of the Fourier 
transform.
However, we need not to use the 
Fourier transform. What is the role 
of the Schr\"odinger equation in the above construction of $\mathcal{H}$?
Note, that in general 
\begin{displaymath}
\Vert \psi \Vert^{2} \equiv
\int \psi^{*}(\vec{x},0)\psi(\vec{x},0) \, {\ud}^{3}{x} = (\varphi,\varphi) = 
\end{displaymath}

\begin{displaymath}
=\int \psi^{*}(\vec{x},t)\psi(\vec{x},t) \, {\ud}^{3}{x}.
\end{displaymath}
This is in accordance with the Born interpretation of $\psi$. Namely, if 
$\psi^{*}\psi(\vec{x},t)$ is the
probability density, then
\begin{displaymath}
\int \psi^{*}\psi \, {\ud}^{3}{x}
\end{displaymath}
has to be preserved in time. In the above construction the Hilbert space 
$\mathcal{H}$ is isomorphic to the 
space of square integrable functions 
$\varphi(\vec{x})\equiv \psi(\vec{x},0)$ -- the set of square integrable
space of initial data for the Schr\"odinger equation, see e.g. \cite{Giulini}. 
The connection between $\psi$ and
$\varphi$ is given by the time evolution 
$U(0,t)$ operator (by the Schr\"odinger equation):
\begin{displaymath}
U(0,t)\varphi=\psi.
\end{displaymath}
The correspondence between $\varphi$ and $\psi$ has all formal properties such 
as in the above Fourier 
construction. Of course, the initial data for the 
Schr\"odinger equation do not cover the whole Hilbert space 
$\mathcal{H}$ of square integrable functions, but the time evolution given by 
the Schr\"odinger equation can be 
uniquely extended on the whole Hilbert space 
$\mathcal{H}$ by the unitary evolution operator $U$. 

The construction can be applied to the particle in the Newton-Cartan spacetime. 
As we implicitly assumed, 
the wave equation is such that the set of its 
admissible initial data is dense in the space of square integrable functions
(we need it for the uniqueness of the extension). Because of the Born 
interpretation the integral
\begin{displaymath}
\int \psi^{*}\psi \, {\ud}^{3}{x}
\end{displaymath}
has to be preserved in time. Denote the space of the initial square integrable 
data $\varphi$ on the simultaneity
hyperplane $t(X)=t$ by ${\mathcal{H}}_{t}$. 
The evolution is, then, an isometry between ${\mathcal{H}}_{0}$ and 
${\mathcal{H}}_{t}$. But such an isometry has to be a unitary operator, 
and the construction is well defined,
\emph{i.e.} the inner product of two 
states corresponding to the wave functions $\psi_{1}$ and $\psi_{2}$
does not depend on the choice of ${\mathcal{H}}_{t}$. 
Let us mention, that 
the wave equation has 
to be linear in accordance with the Born interpretation 
of $\psi$ (any unitary operator is linear,
so, the time evolution operator is 
linear). The space of wave functions $\psi(\vec{x},t) = U(0,t)\varphi(\vec{x})$
isomorphic to the Hilbert space ${\mathcal{H}}_{0}$ of $\varphi$'s is called in 
the common "jargon" the 
"Schr\"odinger picture".  

However, the connection between $\varphi(\vec{x})$ and $\psi(\vec{x},t)$ is not 
unique in general,
if the wave equation possesses a gauge freedom. Namely, 
consider 
the two states $\varphi_{1}$ and $\varphi_{2}$ and ask the question: 
when the two states are equivalent and by this indistinguishable? The answer is 
as follows: they are equivalent if 
\begin{displaymath}
\vert(\varphi_{1},\varphi)\vert \equiv \Big\vert\int \psi_{1}^{*}(\vec{x},t)
\psi(\vec{x},t) \, {\ud}^{3}{x}\Big\vert =
\vert(\varphi_{2},\varphi)\vert \equiv
\end{displaymath}

\begin{equation}\label{row}
\equiv \Big\vert\int \psi_{2}^{*}(\vec{x},t)\psi(\vec{x},t) \, 
{\ud}^{3}{x}\Big\vert,
\end{equation}
for any state $\varphi$ from $\mathcal{H}$, or for all $\psi=U\varphi$ 
($\psi_{i}$ are defined to be = 
$U(0,t)\varphi_{i}$). Substituting $\varphi_{1}$ 
and then 
$\varphi_{2}$ for $\varphi$ and making use of the Schwarz's inequality 
one gets: $\varphi_{2}=e^{i\alpha}\varphi_{1}$,
where $\alpha$ is any 
constant\footnote{This gives the conception of the ray, introduced to Quantum 
Mechanics by Hermann Weyl
[H. Weyl, {\sl Gruppentheorie und Quantenmechanik}, 
Verlag von S. Hirzel in Leipzig (1928)]: a physical 
state does not correspond uniquely to a normed state $\varphi \in \mathcal{H}$, 
but it is uniquely described by a 
\emph{ray}, two states belong to the same 
ray if they differ by a constant phase factor.}.
The situation for $\psi_{1}$ and $\psi_{2}$ is however different. 
In general the condition (\ref{row}) is fulfilled if 
\begin{displaymath}
\psi_{2}=e^{i\xi(t)}\psi_{1}
\end{displaymath}
and the phase factor can depend on time. Of course it has to be consistent with 
the wave equation, that is, together
with a solution $\psi$ to the wave equation 
the wave function $e^{i\xi(t)}\psi$ also is a solution
to the appropriately 
gauged wave equation. \emph{A priori}
one can not exclude the existence of such 
a consistent time evolution. This is not a new observation, it was 
noticed by John von Neumann\footnote{J. v. Neumann, {\sl Mathematical Principles 
of Quantum Mechanics}, University Press, Princeton (1955). He did not mention 
about the gauge freedom on that occasion. But the gauge freedom 
is necessary for the equivalence of $\psi_{1}$ and $\psi_{2} 
= e^{i\xi(t)}\psi_{1}$.}, but it seems that it has never 
been deeply investigated
(probably because the ordinary nonrelativistic 
Schr\"odinger equation has a gauge symmetry 
with constant $\xi$). 
The space of waves $\psi$ describing the system cannot be reduced in the above 
way to any fixed Hilbert space $\mathcal{H}_{t}$ with a fixed $t$. So, 
the existence of the nontrivial gauge freedom leads to the 

\vspace{1ex}

{\bf Hypothesis}. \emph{The two waves $\psi$ and $e^{i\xi(t)\psi}$ are
quantum-mechanically indistinguishable}.

\vspace{1ex}

Moreover, we are obliged to use the whole Hilbert bundle $\mathcal{R}\triangle 
\mathcal{H}: t \to \mathcal{H}_{t}$ over the time instead of a fixed Hilbert
space $\mathcal{H}_{t}$, with the appropriate cross sections as the waves
$\psi$, see the next section for details. 

Consider now an action $T_{r}$ of a group $G$ in the space of waves $\psi$.    
Before we infer some consequences of the assumption that $G$ is a symmetry 
group we need to state a postulate:

\vspace{1ex}

{\bf Classical-like  postulate}. \emph{The group $G$ is a symmetry 
group if and only if the wave equation is invariant under the transformation
$x' = rx, r\in G$ of independent variables and the transformation $\psi'= 
T_{r}\psi$ of the wave function}.

\vspace{1ex}
 
The above postulate is indeed commonly accepted in Quantum Mechanics even when
the gauge freedom is not excluded. But it is a mere application of the symmetry
definition for a classical field equation applied to the wave equation
without any change. The wave $\psi$, is not any classical quantity, like
the electromagnetic intensity. The above {\bf Hypothesis} is not
true for classical field and we have to be careful in forming the appropriate
postulate for the wave equation compatible withe the {\bf Hypothesis}.   
Namely, the two wave equations differing by a mere gauge are indistinguishable.
We call them \emph{gauge-equivalent}. It is therefore natural to assume

\vspace{1ex}

{\bf Quantum postulate}. \emph{The group $G$ is a symmetry 
group if and only if  the transformation $x' = rx, r\in G$ of independent 
variables and the transformation $\psi'= T_{r}\psi$ of the wave function 
transform the wave equation into a gauge-equivalent one}.

\vspace{1ex}

Note, that not all possibilities admitted by the {\bf Hypothesis}
are included in 
{\bf Classical-like postulate}. 

From the {\bf Classical-like postulate} it follows that $\psi$
as well as $T_{r}\psi$ are solutions to exactly the same wave equation,
in view of the invariance of the equation. Therefore, $\psi$ and 
$T_{r}\psi$ belong to the same "Schr\"odinger picture", so that
\begin{displaymath}
T_{r}T_{s}\psi = e^{i\xi(r,s)}T_{rs}\psi,
\end{displaymath}
with $\xi = \xi(r,s)$ independent of the time $t$! This is in accordance
with the known theorem that

\vspace{1ex}

\begin{twr} If $G$ is a symmetry group, then the phase factor $\xi$ should 
be time independent. \end{twr}

\vspace{1ex}

But if we start from the {\bf Quantum postulate} we obtain 
instead
\begin{equation}\label{rowt}
T_{r}T_{s}\psi=e^{i\xi(r,s,t)}T_{rs}\psi
\end{equation}
and get the

\vspace{1ex}

{\bf Theorem 1'} \, \emph{If $G$ is a symmetry group, then the phase 
factor $\xi = \xi(r,s,t)$ is time-dependent in general}. 

\vspace{1ex}

In this paper we propose to accept {\bf Quantum postulate}, which is 
compatible with the {\bf Hypothesis}, and is more in spirit of Quantum 
Mechanics then the {\bf Classical-like postulate}. It should be noted 
that in the special case when the gauge freedom degenerates to
the constant phase the {\bf Quantum postulate} is equivalent to
to the {\bf Classical-like postulate}. 

Acceptation of the {\bf Quantum postulate} gives a new perspective 
for solving the two very difficult problems:
\begin{enumerate}

\item[(a)]
          generally covariant formulation of Quantum Mechanics,

\item[(b)]
          the troubles in the Quantum Field Theory caused by the gauge freedom.

\end{enumerate}
Moreover, with the help of the {\bf Hypothesis} we can see that both
(a) and (b) are deeply connected. Consider the standard treatment in which
the {\bf Hypothesis} is not taken into account and $\mathcal{H}_{0}$
is assumed to be the Hilbert space of all states and 
{\bf Classical-like postulate} is accepted. Then a troubles arise
if we intend to formulate a representation theory of a covariance group
in contradistinction to a symmetry group.
The troubles have their source in the fact that the covariance group transforms 
solution of the wave equation into a solution of the transformed wave equation, 
but the transformed equation is different in form in comparison to the initial one.
That is $T_{r}\psi$ does not belong to the same "Schr\"odinger picture" as
$\psi$, and $T_{r}$ does not act in the Hilbert space $\mathcal{H}_{0}$ of states. 
In view of the paradigm that any reasonable treating of action of any group in 
Quantum Mechanics reduces to a unitary representation of the group in the Hilbert
space of states, there was no natural way for treating the covariance group.
The difficulty disappear if we start from  the {\bf Hypothesis} and
the {\bf Quantum postulate}. Now the states are the appropriate cross
sections in the bundle $\mathcal{R}\triangle \mathcal{H}$ and $T_{r}$ transforms
unitarily fibre $\mathcal{H}_{t}$ onto a fibre $\mathcal{H}_{r^{-1}t}$ and
acts in the same space $\mathcal{R}\triangle \mathcal{H}$ as the symmetry group.  
Note that the space of states degenerates to a fixed fibre $\mathcal{H}_{0}$ 
over $t=0$ if the gauge freedom degenerates to a constant phase.

Before we explain connection to the problem (b), we should resolve a paradox
and then make a comment concerning the Quantum Field Theory.. 
Namely,a natural question arises: why the phase factor $e^{i\xi}$ in 
(\ref{rowt})
is time independent for the Galilean group (even when
the Galilean group is considered as a covariance group)? 
The explanation of the 
paradox is as follows. The Galilean covariance group $G$ induces the 
representation
$T_{r}$ in the space $\mathcal{R}\triangle \mathcal{H}$ 
and fulfills (\ref{rowt}). But, as we will show later on, 
the structure of $G$ is such that 
there always exists a function $\zeta(r,t)$ continuous in $r$ and 
differentiable in $t$ with the help of which one can define a new equivalent 
representation 
$T'_{r}=e^{i\zeta(r,t)}T_{r}$ fulfilling
\begin{displaymath}
T'_{r}T'_{s}=e^{i\xi(r,s)}T'_{rs} 
\end{displaymath}
with a time independent $\xi$. The representations $T_{r}$ and $T'_{r}$ 
are equivalent because $T'_{r}\psi$ and $T_{r}\psi$ are equivalent for all 
$r$ and $\psi$. However this is not the case in general, when the exponent 
$\xi$ depends on the time and 
this time dependence cannot be eliminated in 
such a way as for the Galilean group. We have such a situation when we try 
to find the most general wave equation for a nonrelativistic quantum particle 
in the Newton-Cartan spacetime. The relevant covariance group in this case
is the Milne group which possesses representations with time dependent $\xi$ not 
equivalent to any
representations with a constant (in time) $\xi$. Moreover, 
the only physical
representations of the Milne group are those with time 
dependent $\xi$. 

We make a general comment concerning the 
relation (\ref{rowt}). There is a physical 
motivation to investigate 
representations $T_{r}$ fulfilling (\ref{rowt}) with $\xi$ depending on 
spacetime point $p$:
\begin{equation}\label{rowX}
T_{r}T_{s}=e^{i\xi(r,s,p)}T_{rs}.
\end{equation}
Namely, in the Quantum Field Theory the spacetime coordinates of 
$p \in \mathcal{M}$ play the role of parameters such 
as the time 
plays in the nonrelativistic theory 
(recall that, for example, the wave functions
 $\psi$ of the Fock space of 
the quantum electromagnetic field are functions of the Fourier components 
of the field,
 the spacetime coordinates playing the role of parameters like 
the time $t$ in the nonrelativistic Quantum Mechanics). 
By this the two wave functions $\psi$ and $\psi' = e^{i\xi(p)}\psi$ are 
indistinguishable in the sense that 
they give the same transition 
probabilities: $\vert(\psi(p),\phi(p))\vert = \vert(\psi'(p),\phi(p))\vert$, 
for any 
$\phi$. In an analogous way we get the Hilbert bundle 
$\mathcal{M}\triangle \mathcal{H}$ over the spacetime $\mathcal{M}$ and 
appropriate cross sections as the wave functions $\psi$, see the next section
for details. 

Now, we return to the problem (b).
It should be mentioned at this place that the troubles in QFT 
generated
by the gauge freedom are of general character, and are well known. 
For example, there do not exist vector particles with 
helicity = 1, which is a consequence
of the theory of unitary 
representations of the Poincar\'e group, as was shown by Jan {\L}opusza\'nski 
\cite{Lopuszanski}. This is apparently in contradiction with the existence of 
vector particles with
helicity = 1 in nature -- the photon, which is connected 
with the electromagnetic four-vector potential.
The connection of the problem 
with the gauge freedom is 
well known \cite{Lopuszanski}. We omit however the 
difficulty if we allow the inner product in
the "Hilbert space" to be not 
positively defined, see  \cite{Gupta}, 
or  \cite{BRST}.
Due to \cite{Lopuszanski}, the vector potential (promoted to be an 
operator valued distribution in QED)
cannot be a vector field, if one 
wants to have the inner product positively defined -- together with
the coordinate transformation the gauge transformation has to be applied, 
which breaks the vector
character of the potential. Practically it means that 
any gauge condition which brings the theory into
the canonical form such that 
the quantization procedure can be consequently applied (with the positively
defined inner product in the Hilbert space) breaks the four-vector character 
of the electromagnetic potential, the 
Coulomb gauge condition is an example. 
To achieve the Poincar\'e symmetry of Maxwell
equations with such a gauge 
condition (the Coulomb gauge condition for example), it is impossible to preserve 
the vector character 
of the potential -- together with the coordinate 
transformation a well defined (by the coordinate transformation)
gauge 
transformation $f$ has to be applied:
\begin{displaymath}
A_{\mu} \to A'_{\mu'} = \frac{\partial x^{\nu}}{\partial x^{\mu'}}(A_{\nu} 
+ \partial_{\nu}f).
\end{displaymath}
This means that the electromagnetic potential can form
a ray representation $T_{r}$ (in the sense of (\ref{rowX})) of the Poincar\'e 
group at most, with
 the spacetime-dependent factor $e^{i\xi}$ if the scalar 
product is positively defined. 
One may ask: how possible is it if the 
Poincar\'e
 group is not only a covariance group but at the same time a 
symmetry group? The solution of this paradox
 on the grounds of the 
existing theory is rather obscure. We propose the following solution.
The Theorem 1 is true for the symmetry group but under the assumption 
that the fundamental space describing the states of a quantum system is 
 the ordinary Hilbert space and the {\bf Classical-like postulate} is true.
But we have presented serious objections to this assumption. 
Namely, the nonrelativistic quantum theory can be 
reconstructed from the more general assumption about the space of quantum 
mechanical states saying that it compose the space of appropriate cross sections
of the Hilbert bundle $\mathcal{R}\triangle\mathcal{H}$ 
over time $t \in \mathcal{R}$. The Schr\"odinger equation can be uniquely 
reconstructed from the generalized ray representations of the Galilean group. 
We watch for also a more fundamental justification of this assumption in the 
presumption that the time is a purely classical variable in the nonrelativistic 
quantum mechanics or so to speak a parameter. The most general unitary
representation of the locally compact commutative group of the time 
real line acts in a Hilbert bundle $\mathcal{R}\triangle\mathcal{H}$  
over the time, see Mackey \cite{Mackey}. So, the assumption about the 
"classicity" of the time $t$ fixes the structure of space of quantum states 
to be a subset of cross sections of a Hilbert bundle over the time.  
This is the peculiar property of the Galilean group 
structure that the whole construction degenerates as if we were started from
the ordinary ray representation over the ordinary Hilbert space and
the Theorem 1 is true in this case, but only accidentally.  
The generalization to the relativistic case is natural. First we postulate
the spacetime coordinates to be classical commutative variables, which 
leads to the Hilbert bundle $\mathcal{M}\triangle\mathcal{H}$  over the 
space-time manifold $\mathcal{M}$. The factor of the representation of the Poincar\'e 
group acting in the bundle $\mathcal{M}\triangle\mathcal{H}$ has not to be 
a constant with respect to space-time coordinates even when it is a symmetry 
group. 

To realize the above program consequently  
we are forced to generalize
 the Bargmann's theory of factors to embrace the 
spacetime-dependent factors of representations acting in a Hilbert
bundle over the space-time (time).

\section{Generalized Wave Rays and Operator Rays}\label{generalization}

In this section we give the strict mathematical definitions to the
notions of the preceding section and formulate the problem
stated there in the strict way. From the pure mathematical point 
of view the analysis of spacetime dependent $\xi(r,s,p)$ is 
more general, 
so we confine ourselves to this case at the outset, but we mark the place at 
which important 
difference arises between the two cases\footnote{It becomes 
clear in the further analysis that the group $G$ in question has 
to fulfill the consistency condition that for 
any $r\in G$, $rt$ is a 
function of time only in the case of the nonrelativistic theory with 
(\ref{rowt}).}. 

\vspace{1ex}

Let us remind some definitions, compare e.g. \cite{Mackey}. 
Let $\mathcal{M}$ be a set and $G$ be a group. 
A function $p,g \to pg$ from $\mathcal{M}\times G$ to $G$ 
will be said to convert the set $\mathcal{M}$ into a $G$-space if the 
following two conditions are satisfied
\begin{enumerate}

\item[(a)]
$g_{2}(g_{1}p) = (g_{2}g_{1})p$, for all $p \in \mathcal{M}$,
$g_{1},g_{2}\in G$.

\item[(b)]
 $ep=p$, for all $p\in\mathcal{M}$, where $e$ is the identity of $G$.
  
\end{enumerate}
We say that $G$ acts on the left. If we write $pg$ and assume
$(pg_{1})g_{2}=p(g_{1}g_{2})$ instead, we say that $G$ acts on the right. 
If the function $\mathcal{M} \times G \to G$ is smooth then we say
that $G$ acts smoothly in $\mathcal{M}$.

\vspace{1ex}
 
By a \emph{Hilbert bundle over} $\mathcal{M}$ or a \emph{Hilbert bundle
with base} $\mathcal{M}$ we shall mean an assignment 
$\mathcal{H}: p \to\mathcal{H}_{p}$ of a Hilbert space $\mathcal{H}_{p}$
to each $p \in \mathcal{M}$. The set of all pairs $(p,\psi)$ with
$\psi \in \mathcal{H}_{p}$ will be denoted by 
$\mathcal{M}\triangle\mathcal{H}$ and called the \emph{space of the bundle}.  
By a \emph{cross section} of our bundle we shall mean an assignment
$\psi: p \to \psi_{p}$ of a member of $\mathcal{H}_{p}$ to each 
$p\in\mathcal{M}$. If $\psi$ is a cross section and $(p_{0},\phi_{0})$
a point of $\mathcal{M}\triangle\mathcal{H}$, we may form a scalar product
$(\phi_{0},\psi_{p_{0}})$. In this way every cross section $\psi$ defines a 
complex-valued function $f_{\psi}$ on $\mathcal{M}\triangle\mathcal{H}$. 
By a \emph{Borel Hilbert bundle} we shall mean
a Hilbert bundle  together with an analytic Borel structure in
$\mathcal{M}\triangle\mathcal{H}$ such that the following conditions 
are fulfilled
\begin{enumerate}
\item[(1)]  Let $\pi(p,\psi)=p$. Then $E \subseteq \mathcal{M}$ is a Borel 
set if and only if $\pi^{-1}(E)$ is a Borel set in 
$\mathcal{M}\triangle\mathcal{H}$.

\item[(2)]  There exist countably many cross sections 
$\psi^{1}, \psi^{2}, \ldots$ such that 
\begin{itemize}
\item[(a)] the corresponding complex-valued functions on 
$\mathcal{M}\triangle\mathcal{H}$ are Borel functions,

\item[(b)] these Borel functions separate points in the sense 
that no two distinct points $(p_{i},\phi_{i})$ of 
$\mathcal{M}\triangle\mathcal{H}$ assign the same values to all
$\psi^{j}$ unless $\phi_{1}=\phi_{2}=0$, and

\item[(c)] $p \to (\psi^{i}(p),\psi^{j}(p))$ is a Borel function for all 
$i$ and $j$.
\end{itemize}
\end{enumerate}

A cross section is said to be a \emph{Borel cross section} 
if the function on $\mathcal{M}\triangle\mathcal{H}$ defined by the 
cross section is a Borel function. All Borel cross sections
compose a linear space under the obvious operations, see \cite{Mackey}.
Now let $\mu$ be a measure on $\mathcal{M}$. The cross section
$p \to \varphi_{p}$ is said to be \emph{square summable} with respect
to $\mu$ if 
\begin{displaymath}
\int_{\mathcal{M}} (\varphi_{p}, \varphi_{p}) \, d\mu(p) < \infty.
\end{displaymath} 
The space $\mathcal{L}^{2}(\mathcal{M}, \mu, \mathcal{H})$  of 
all equivalence classes of square summable cross sections, where two
cross sections $\varphi$ and $\varphi'$ are in the same equivalence 
class if $\varphi_{p}=\varphi'_{p}$ for almost all $p \in \mathcal{M}$,
forms a Hilbert space with the inner product given by
\begin{displaymath}
(\varphi, \theta) = \int_{\mathcal{M}} (\varphi_{p}, \theta_{p}) \, d\mu(p),
\end{displaymath}
see \cite{Mackey}. It is called the direct integral of the $\mathcal{H}_{p}$
with respect to $\mu$ and is denoted by 
$\int_{\mathcal{M}} \mathcal{H}_{p} \, d\mu(p)$. 

\vspace{1ex}

The identification with the previous section is partially suggested by the 
notation. We make the identification more explicit. The set $\mathcal{M}$
plays the role of the spacetime or the real line $\mathcal{R}$  of the time 
$t$ respectively. The unitary representation of 
the commutative group of coordinates in the spacetime $\mathcal{M}$ acts 
precisely in the distinguished subset $\mathcal{L}^{2}(\mathcal{M}, \mu, 
\mathcal{H})$ of all Borel cross sections. We refer the reader to 
\cite{Mackey} and literature therein for a detailed description of this 
representation. The wave functions  $\psi$ of 
the preceding section are the Borel cross sections of 
$\mathcal{M}\triangle\mathcal{H}$ but they do not belong to the  
subset $\mathcal{L}^{2}(\mathcal{M},\mu,\mathcal{H})$ of cross sections 
which are square integrable. Rather the separate Hilbert spaces 
$\mathcal{H}_{p}$ with their inner products play a role in experiments 
than the inner product in the direct integral product of them. We have also 
used interchangeably $\psi(p)$ and $\psi_{p}$ as well as 
$(\psi_{p}, \theta_{p})$ and $(\psi, \theta)_{p}$. 

The physical interpretation ascribed to the cross section $\psi$ 
is as follows. Each experiment is, out of its very nature, a 
spatiotemporal event. To each act of measurement carried out at the 
spacetime point $p_{0}$ we ascribe a self-adjoint operator $Q_{p_{0}}$ acting 
in the Hilbert space $\mathcal{H}_{p_{0}}$ and ascribe to the spectral 
theorem for $Q_{p_{0}}$ the standard interpretation. Hence, assuming for 
simplicity that $Q_{p_{0}}$ is bounded, if $\phi_{0} \in \mathcal{H}_{p_{0}}$ 
and $\lambda_{0} =\lambda_{o}(p_{0})$ is a characteristic vector and its 
corresponding characteristic value of $Q_{p_{0}}$  respectively then we have 
the following statement.  
\emph{If the experiment corresponding to} $Q_{p}$  
\emph{was performed at the spatiotemporal event} $p_{0}$
\emph{on a system in the state described
by the cross section $\psi$, then the probability of the measurement
value to be $\lambda_{0}(p_{0})$ and the system to be found
in the state described by $\phi$ such that $\phi(p_{o})=\phi_{0}$
after the experiment is given by the square of the absolute value of 
the Borel function $\vert f_{\psi}(p_{0},\phi_{0}) \vert^{2} 
= \vert (\phi_{0},\psi_{p_{0}})\vert^{2}$  
induced by the cross section $\psi$}. In the nonrelativistic 
case the above statement is a mere rephrasing of the well
established knowledge. 

\vspace{1ex}

By an \emph{isomorphism} of the Hilbert bundle 
$\mathcal{M}\triangle\mathcal{H}$ with the Hilbert bundle
$\mathcal{M}'\triangle\mathcal{H}'$ we shall mean a Borel isomorphism $T$
of $\mathcal{M}\triangle\mathcal{H}$ on $\mathcal{M}'\triangle\mathcal{H}'$ 
such that for each $p \in \mathcal{M}$ the restriction of $T$ to
$p \times \mathcal{H}_{p}$ has some $q \times \mathcal{H}_{q}'$
for its range and is unitary when regarded as a map of $\mathcal{H}_{p}$
on $\mathcal{H}_{q}'$ The induced map carrying $p$ into $q$ is clearly
a Borel isomorphism of $\mathcal{M}$ with $\mathcal{M}'$ and we denote it
by $T^{\pi}$. The above defined $T$ is said to be an \emph{automorphism}
if $\mathcal{M}\triangle\mathcal{H}=\mathcal{M}'\triangle\mathcal{H}'$. 
Note that for any automorphism $T$ we have 
$(T\psi,T\phi)_{T^{\pi}p}=(\psi,\phi)_{p}$, but in general
$(T\psi,T\phi)_{p}\neq(\psi,\phi)_{p}$. By this any automorphism $T$ is
what is frequently called a \emph{bundle isometry}. 

The function $r \to T_{r}$ from a group $G$ into the set of automorphisms
(bundle isometries) of $\mathcal{M}\triangle\mathcal{H}$ is said to be
a \emph{general factor representation} of $G$ associated 
to the action $G \times \mathcal{M} \ni r,p \to r^{-1}p \in \mathcal{M}$ of
$G$ on $\mathcal{M}$ if $T_{r}^{\pi}(p) \equiv r^{-1}p$ for all $r \in G$,
and $T_{r}$ satisfy the condition (\ref{rowX}). 

\vspace{1ex}

Of course $T_{r}$ is to be identified with the one of the preceding
section. Our further specializing assumptions partly following from the above
interpretation are as follows. We assume $\mathcal{M}$ to be endowed with 
the manifold structure inducing a topology associated with the above
assumed Borel structure. We confine ourselves to a finite dimensional
Lie group $G$ which acts smoothly and transitively on the spacetime 
$\mathcal{M}$, so, a $G$-invariant measure $\mu$ exists on $\mathcal{M}$. 

\vspace{1ex}

By a \emph{factor representation} of a Lie group we mean a general
factor representation with the exponent $\xi(r,s,p)$ differentiable in 
$p \in \mathcal{M}$.

Now we define the \emph{operator ray} $\boldsymbol{T}$ corresponding
to a given bundle isometry operator $T$ to be set of operators 
\begin{displaymath}
{\boldsymbol{T}}=\{\tau T ,  p \to \tau(p) \in {\mathcal{D}} \, \, 
\textrm{and} \, \,  \vert\tau \vert = 1\}
,
\end{displaymath} 
where $\mathcal{D}$ denotes the set of all differentiable 
real functions on $\mathcal{M}$.  
Any $T\in {\boldsymbol{T}}$ will be called a \emph{representative} of the 
ray ${\boldsymbol{T}}$. 
The product ${\boldsymbol{TV}}$ is defined as the 
set of all products $TV$ such that
 $T\in {\boldsymbol{T}}$ and 
$V \in {\boldsymbol{V}}$. 

\vspace{1ex}

Note that not all Borel sections are physically realizable.
Interpreting the discussion of the preceding section in
the Hilbert bundle language we see that the role of the 
Schr\"odinger equation is essentially to establish all
the physical sections. Any two sections $\psi(p)$ and
$\psi'(p) =e^{i\zeta(p)}\psi(p)$ are indistinguishable
giving the same probabilities $\vert f_{\psi}\vert^{2}
=\vert f_{\psi'}\vert^{2}$. After this any group
$G$ acting in $\mathcal{M}$ induces a \emph{ray representation} of $G$,
i.e. a mapping $r \to \boldsymbol{T}_{r}$ of $G$ into the space of rays of
bundle automorphisms (bundle isometries) of 
$\mathcal{M}\triangle\mathcal{H}$, fulfilling the condition
\begin{displaymath}
\boldsymbol{T}_{r}\boldsymbol{T}_{s}=\boldsymbol{T}_{rs}.
\end{displaymath}
To any cross section $\psi$ we define the corresponding 
\emph{ray} $\boldsymbol{\psi} =\{e^{i\zeta(p)}\psi(p), 
\zeta \in \mathcal{D}\}$. if $\psi$ is a physical cross section then we get 
the \emph{physical ray} of the preceding section. 
Selecting a representative $T_{r}$ for each $\boldsymbol{T}_{r}$ we get
a factor representation fulfilling (\ref{rowX}).
Note that $T_{r}$ transforms rays 
into rays, and we have $T_{r}(e^{i\xi(p)}\psi)=e^{i\xi_{r}(p)}T_{r}\psi$. 
In the sequel we assume that the the operators
$T_{r}$ are such that   $\xi_{r}(p)=\xi(r^{-1}p)$, where $r^{-1}p$ denote 
the the action of $r^{-1} \in G$ on the spacetime
point $p \in \mathcal{M}$. Note again, that this is a natural assumption 
which takes place in practice.    

Now we make the last assumption, namely the assumption that all transition 
probabilities vary continuously with
the continuous variation of the coordinate transformation $s \in G$:
\begin{enumerate}
\item[] For any element $r$ in $G$, any ray ${\boldsymbol{\psi}}$ 
and any positive $\epsilon$ there exists a neighborhood $\mathfrak{N}$ of $r$ 
on $G$ such that 
$d_{p}(\boldsymbol{T}_{s}\boldsymbol{\psi},
\boldsymbol{T}_{r}\boldsymbol{\psi}) < \epsilon$ if 
$s \in \mathfrak{N}$
 and $p \in \mathcal{M}$,
\end{enumerate}
where
\begin{displaymath}
d_{p}(\boldsymbol{\psi_{1}},\boldsymbol{\psi_{2}}) = 
\inf_{\psi_{i} \in \boldsymbol{\psi}_{i}}
\Vert \psi_{1} - \psi_{2} \Vert_{p} 
= \sqrt{2\vert1-\vert(\psi_{1},\psi_{2})_{p}\vert \, \vert}.
\end{displaymath}

Basing on the continuity assumption one can prove the following 

\vspace{1ex}

\begin{twr} Let ${\boldsymbol{T}}_{r}$ be a continuous ray representation of 
a group $G$. For all 
$r$ in a suitably chosen neighborhood $\mathfrak{N}_{0}$ 
of the unit element $e$ of $G$ one may select 
a strongly continuous set of 
representatives $T_{r}\in {\boldsymbol{T}}_{r}$. That is, for any compact set 
$\mathcal{C} \subset \mathcal{M}$, any wave function $\psi$, 
any $r \in {\mathfrak{N}}_{0}$ 
and any positive $\epsilon$ there exists a 
neighborhood $\mathfrak{N}$ of $r$ such that $\Vert T_{s}\psi 
- T_{r}\psi \Vert_{p} < \epsilon$ if $s \in \mathfrak{N}$ and 
$p \in \mathcal{C}$. \end{twr}     

\vspace{1ex}

There are many possible selections of such factor representations.
But many among them differ by a mere differentiable
phase factor and are physically indistinguishable. 
We call them to be equivalent. Our task is to classify all 
possible factor representations with respect 
to this equivalence.

\section{Local Exponents}\label{exponents}

The representatives $T_{r} \in \boldsymbol{T}_{r}$ selected as in the Theorem
 2 will be called \emph{admissible}
 and the representation $T_{r}$ obtained 
in this way the \emph{admissible} representation. 
There are infinitely many 
possibilities of such a selection of admissible representation 
$T_{r}$. We confine ourselves to the local \emph{admissible} representations 
defined on a fixed neighborhood $\mathfrak{N}_{o}$ of $e \in G$, as in the 
Theorem 2. 

Let $T_{r}$ be an \emph{admissible} representation. With the help of the 
phase
 $e^{i\zeta(r,p)}$ with a real function $\zeta(r,p)$ differentiable in 
$p$ and continuous in $r$ we can define
\begin{equation}\label{4}
T'_{r} = e^{i\zeta(r,p)}T_{r},
\end{equation}
which is a new \emph{admissible} representation. This is trivial, if one 
defines in the appropriate way 
the continuity of $\zeta(r,p)$ in $r$. Namely, 
from the Theorem 2 it follows that the continuity has to be defined
in the following way. The function $\zeta(r,p)$ \emph{will be called strongly 
continuous in r at $r_{0}$ if and 
only if for any compact set $\mathcal{C} 
\subset \mathcal{M}$
 and any positive $\epsilon$ there exist a 
neighborhood ${\mathfrak{N}}_{0}$ of $r_{0}$ such that } 
\begin{displaymath} 
\vert \zeta(r_{0},p) - \zeta(r,p) \vert < \epsilon,
\end{displaymath}
\emph{for all} $r \in {\mathfrak{N}}_{0}$ \emph{and for all} 
$p \in \mathcal{C}$. 
 But the converse is also true. Indeed,
if $T'_{r}$ also is an \emph{admissible} representation, then (\ref{4}) 
has to be fulfilled for a real function $\zeta(r,p)$
differentiable in $p$ because $T'_{r}$ and $T_{r}$ belong to the same ray, 
and moreover, because both 
$T'_{r}\psi$ and $T_{r}\psi$ are strongly 
continuous (in $r$ for any $\psi$) then $\zeta(r,p)$ has to be 
\emph{strongly continuous} (in $r$).

Let $T_{r}$ be an \emph{admissible} representation, and by this continuous 
in the sense indicated in the 
Theorem 2. One can always choose the above 
$\zeta$ in such a way that $T_{e} = 1$ as will be assumed 
in the sequel. 

Because $T_{r}T_{s}$ and $T_{rs}$ belong to the same ray one has
\begin{equation}\label{5}
T_{r}T_{s} = e^{i\xi(r,s,p)} T_{rs}
\end{equation}    
with a real  function $\xi(r,s,p)$  differentiable in $p$. From the fact that 
$T_{e} = 1$ we have
\begin{equation}\label{9}
\xi(e,e,p) = 0.
\end{equation}  
From the associative law $(T_{r}T_{s})T_{g} = T_{r}(T_{s}T_{g})$ one gets
\begin{equation}\label{10}
\xi(r,s,p) + \xi(rs,g,p) = \xi(s,g,r^{-1}p) + \xi(r,sg,p).
\end{equation} 
The formula (\ref{10}) is very important and our analysis largely rests on 
this relation.
 From the fact that the representation $T_{r}$ is 
\emph{admissible} follows that 
the exponent $\xi(r,s,p)$ is continuous in 
$r$ and $s$. Indeed, take a $\psi$ belonging
to a unit ray $\boldsymbol{\psi}$, then making use of (\ref{5}) we get
\begin{displaymath}
e^{i\xi(r,s,p)}(T_{rs} - T_{r's'})\psi + (T_{r'}(T_{s'} - T_{s})\psi + 
(T_{r'} - T_{r})T_{s}\psi 
\end{displaymath}  

\begin{displaymath}
= (e^{i\xi(r',s',p)} - e^{i\xi(r,s,p)}) T_{r's'}\psi.
\end{displaymath}
Taking norms $\Vert \centerdot \Vert_{p}$ of both sides, we get 
\begin{displaymath}
\vert e^{i\xi(r',s',p)} - e^{i\xi(r,s,p)} \vert \leq 
\Vert (T_{r's'} - T_{rs})\psi \Vert_{p} + 
\end{displaymath}

\begin{displaymath}
+ \Vert T_{r'}(T_{s'} - T_{s})\psi \Vert_{p} + \Vert (T_{r'} - 
T_{r})T_{s}\psi \Vert_{p}.
\end{displaymath} 
From this inequality and the continuity of $T_{r}\psi$, the continuity of 
$\xi(r,s,p)$ in $r$ and $s$ follows.
Moreover, from the Theorem 2 and the above inequality the 
\emph{strong continuity} of 
$\xi(r,s,p)$ in $r$ and $s$ follows.  

The formula (\ref{4}) suggests the following definition. Two 
\emph{admissible} representations
$T_{r}$ and $T'_{r}$ are called \emph{equivalent} if and only if 
$T'_{r} = e^{i\zeta(r,p)}T_{r}$ 
for some real function $\zeta(r,p)$ 
differentiable in $p$ and \emph{strongly continuous} in $r$. So, making use 
of 
(\ref{5}) we get $T'_{r}T'_{s} = e^{i\xi'(r,s,p)}T'_{rs}$, where
\begin{equation}\label{13}
\xi'(r,s,p) = \xi(r,s,p) + \zeta(r,p) + \zeta(s,r^{-1}p) - \zeta(rs,p).
\end{equation}
Then the two exponents $\xi$ and $\xi'$ are equivalent if and only if 
(\ref{13}) is fulfilled with
 $\zeta(r,p)$ \emph{strongly continuous} in 
$r$ and differentiable in $p$. 

From (\ref{9}) and (\ref{10}) immediately follows that
\begin{equation}\label{11}
\xi(r,e,p) = 0 \, \, \, and \, \, \, \xi(e,g,p) = 0,
\end{equation}

\begin{equation}\label{12}
\xi(r,r^{-1},p) = \xi(r^{-1},r,r^{-1}p).
\end{equation}
The relation (\ref{13}) between $\xi$ and $\xi'$ will be written in short by 
\begin{equation}\label{13'}
\xi' = \xi + \Delta[\zeta].
\end{equation}
The relation (\ref{13}) between exponents $\xi$ and $\xi'$ is reflexive, 
symmetric and transitive. Indeed, 
we have: $\xi = \xi + \Delta[\zeta]$ with 
$\zeta =0$. Moreover, if $\xi' = \xi + \Delta[\zeta]$ then 
$\xi = \xi' + \Delta[-\xi]$. At last if $\xi' = \xi +\Delta[\zeta]$ and 
$\xi'' = \xi' + \Delta[\zeta']$, then 
$\xi''= \xi + \Delta[\zeta + \zeta']$. 
So the relation is an equivalence relation, and will be sometimes
denoted by $\xi' \equiv \xi$. The equivalence relation preserves the linear 
structure, that is 
if $\xi_{i} \equiv \xi'_{i}$ (with the appropriate 
$\zeta_{i}$-s) then $\lambda_{1} \xi_{1} + \lambda_{2}\xi_{2}
\equiv \lambda_{1}\xi'_{1} + \lambda_{2}\xi'_{2}$ (with 
$\zeta = \lambda_{1}\zeta_{1} + \lambda_{2}\zeta_{2}$).

We introduce now the group $H$, the very important notion for the further 
investigations. It is evident
 that all operators $T_{r}$ contained in all 
rays $\boldsymbol{T}_{r}$ form a group under multiplication.
Indeed, consider an \emph{admissible} representation $T_{r}$ with a well 
defined $\xi(r,s,p)$ in the formula (\ref{5}).
Because any $T_{r} \in \boldsymbol{T}_{r}$ has the form $e^{i\theta(p)}T_{r}$ 
(with a real and differentiable
 $\theta$), one has
\begin{equation}\label{H-action}
\Big( e^{i\theta(p)}T_{r}\Big) \Big( e^{i\theta'(p)}T_{s}\Big) = 
e^{i\{\theta(p) + \theta'(r^{-1}p) + \xi(r,s,p)\}}T_{rs}.
\end{equation}
This important relation suggest the following definition of the local group 
$H$ connected with the \emph{admissible}
 representation or with the exponent 
$\xi(r,s,p)$.
 Namely, $H$ consists of the 
pairs $\{\theta(p), r\}$ where 
$\theta(p)$ is a differentiable real function and $r \in G$. The 
multiplication
 rule, suggested by the above relation, is defined as follows
\begin{equation}\label{15}
\{\theta(p),r\} \centerdot \{\theta'(p),r'\} = \{\theta(p) + 
\theta'(r^{-1}p) + \xi(r,r',p), \, rr' \}.
\end{equation}
The associative law for this multiplication rule is equivalent to (\ref{10}) 
(in a complete
 analogy with the classical Bargmann's theory). The pair 
$\check{e} =\{0,e\}$ plays the role of the unit element in $H$. For any 
element $\{\theta(p), r\} \in H$ there 
exists the inverse 
$\{\theta(p),r\}^{-1} = \{-\theta(rp) - \xi(r,r^{-1},rp), \, r^{-1} \}$. 
Indeed, from (\ref{12}) it follows that 
$\{\theta, r\}^{-1} \centerdot 
\{\theta,r \} = \{\theta, r\} \centerdot \{\theta, r\}^{-1} = \check{e}$. 
The elements
 $\{\theta(p), e\}$ form an abelian subgroup $N$ of $H$. Any 
$\{\theta,r\} \in H$ can be uniquely written as
 $\{\theta(p),r\} = 
\{\theta(p),e\} \centerdot \{ 0, r\}$. Also the same element can be uniquely 
expressed in the form 
$\{\theta(p),r\} = \{0,r\} \centerdot \{\theta(rp),e\}$. 
So, we have $H = N \centerdot G = G \centerdot T$. 
The abelian subgroup $N$ 
is a normal factor subgroup of $H$. But this time $G$ does not form any 
normal 
factor subgroup of $H$ (contrary to the classical case investigated 
by Bargmann, when the exponents do not
 depend on $p$). So, this time $H$ is 
not direct product $N \otimes G$ of $N$ and $G$, but it is a 
semidirect 
product $N \circledS G$ of $N$ and $G$, see e.g. \cite{Nachbin} where the 
semi-direct 
product of two continuous groups is investigated in detail. 
In 
this case however the theorem that $G$ 
is locally isomorphic
 to the factor group $H/N$ is still valid, see 
\cite{Nachbin}. Then the group $H$ composes a
 \emph{semicentral extension} 
of $G$ and not a central extension of $G$ as in the Bargmann's theory.

The rest of this paper is based on the following reasoning (the author 
was largely
 inspired by the Bargmann's work \cite{Bar}). 
If the two exponents 
$\xi$ and $\xi'$ are \emph{equivalent},
 that is $\xi' = \xi + \Delta[\zeta]$, 
then the \emph{semicentral extensions} $H$ and $H'$ connected
with $\xi$ and $\xi'$ are homomorphic. The homomorphism $h: \{\theta,r\} 
\mapsto \{\theta', r'\}$
 is given by 
\begin{equation}\label{izo}
\theta'(p) = \theta(p) - \zeta(r,p), \, \, r' = r.
\end{equation}
 Using an \emph{Iwasawa-type
 construction} we show that any exponent 
$\xi(r,s,p)$ is equivalent to a differentiable one
 (in $r$ and $s$). 
We can confine ourselves then to the differentiable $\xi$ and $\xi'$. 
We show  that  $\zeta(r,p)$ is also differentiable function of $(r,p)$. 
Moreover, we show that any $\xi$ is equivalent to the canonical one, 
that is such $\xi$ which is differentiable and for which 
$\xi(r,s,p) = 0$ whenever $r$ and $s$ 
belong to the same one-parameter subgroup.
Then we can restrict the investigation to the canonical $\xi$  
and we consider the subgroup of all elements $\{\theta(p),r\} \in H$ with 
differentiable $\theta(p)$. Let us denote the subgroup by the same symbol $H$ 
for simplicity. We embed the subgroup in an infinite dimensional Lie group 
$D$ with manifold structure modeled on a Banach space. Then we consider the 
subgroup $\overline{H}$ which is a closure of $H$ in $D$. After this
$\overline{H}$ turns into a Lie group and the homomorphism (\ref{izo})
becomes to be an isomorphism of the two Lie groups.  So, the group 
$\overline{H}$ has the Banach Lie algebra $\overline{\mathfrak{H}}$.   
We apply the general theory of analytic groups developed by 
\cite{Birkhoff} and \cite{Dynkin}. From this theory follows that the 
correspondence between the local $\overline{H}$ and $\overline{\mathfrak{H}}$ 
is bi-unique and one can construct uniquely the local group $\overline{H}$
from the algebra $\overline{\mathfrak{H}}$ also. As we will see 
the algebra defines a spacetime dependent antilinear form 
$\Xi$ on the Lie algebra
 $\mathfrak{G}$ of $G$, the so called 
\emph{infinitesimal exponent} $\Xi$. By this we reduce the classification
of local $\xi$'s which define $\overline{H}$'s to the classification 
of $\Xi$'s which define $\overline{\mathfrak{H}}$'s. So, we will simplify 
the 
problem of the classification of local $\xi$'s to a largely linear 
problem.

\section{Local Exponents of Lie Groups}

{\bf Iwasawa construction}. Denote by ${\ud}r$ and ${\ud}^{*}r$ the left and 
right invariant Haar
 measure on $G$. Let $\nu(r)$ and $\nu^{*}(r)$ be two 
infinitely differentiable
 functions on $G$ with compact supports contained 
in the fixed neighborhood $\mathfrak{N}_{0}$ 
of $e$. Multiplying them by the 
appropriate constants we can always reach:
 $\int_{G} \nu(r) \, {\ud}r = 
\int_{G} \nu^{*}(r) \, {\ud}^{*}r =1$. Let $\xi(r,s,p)$ be any
\emph{admissible} local exponent defined on $\mathfrak{N}_{0}$. We will 
construct a
 differentiable (in $r$ and $s$) exponent $\xi''(r,s,p)$ which is 
\emph{equivalent} to $\xi(r,s,p)$ and is defined
 on $\mathfrak{N}_{0}$, in 
the following two steps:
\begin{displaymath}
\xi' = \xi + \Delta[\zeta], \, \, \textrm{with} \, \, \zeta(r,p) 
= - \int_{G} \xi(r,l,p) \nu(l) \, {\ud}l, 
\end{displaymath}

\begin{displaymath}
\xi'' = \xi' + \Delta[\zeta'], \, \, \textrm{with} \, \, \zeta'(r,p) 
= - \int_{G} \xi'(u,r,up)\nu^{*}(u) \, {\ud}^{*}u.
\end{displaymath}
A rather simple computation in which we use (\ref{13}) and (\ref{10}) 
and the invariance property of the Haar measures gives:
\begin{displaymath}
\xi''(r,s,p) = \int\!\!\int_{G} \xi(u,l,ur^{-1}p) \{\nu(s^{-1}l) - \nu(l) \}
\{ \nu^{*}(ur^{-1}) - \nu^{*}(u) \} \, {\ud}l \, {\ud}^{*}u.
\end{displaymath}
Because $\nu$ and $\nu^{*}$ are differentiable (up to any order) and 
$\xi(r,s,p)$ is a differentiable function of $p \in \mathcal{M}$ 
(up to any order) then $\xi''(r,s,p)$ is a differentiable 
(up to any order)
exponent in all variables $(r,s,p)$.

\vspace{1ex}

\begin{lem} If two differentiable exponents $\xi$ and $\xi'$ are 
equivalent,
that is, if $\xi' = \xi + \Delta[\zeta]$, then 
$\zeta(r,p)$ is differentiable in $r$.\end{lem}

\vspace{1ex}

\emph{Proof}. Clearly, the function $\chi(r,s,p) = \xi'(r,s,p) 
- \xi(r,s,p)$ is differentiable. Similarly the function
$\eta(r,p) = \int_{G} \chi(r,u,p) \nu(u) \, {\ud}u$, 
where $\nu$ is defined as in the Iwasawa construction,
is a differentiable function. But the difference $\zeta' 
= \eta - \zeta$ is equal 
\begin{displaymath}
\zeta'(r,p) = \int_{G} \{\zeta(u,r^{-1}p) 
- \zeta(ru,p) \} \nu(u) \, {\ud}u =
\end{displaymath}   

\begin{displaymath}
= \int_{G} \{ \zeta(u,r^{-1}p)\nu(u) 
-\zeta(u,p)\nu(r^{-1}u) \} \, {\ud}u 
\end{displaymath}
and clearly it is a differentiable function. By this $\zeta 
= \eta - \zeta'$ also is a differentiable function
(recall that $\zeta(r,p)$ is differentiable function of 
$p \in \mathcal{M}$).

\vspace{1ex}

\begin{lem} Every (local) exponent of one-parameter group is 
equivalent to zero. \end{lem}

\vspace{1ex}

\emph{Proof}. We can map such a group $r = r(\tau) \rightleftarrows 
\tau$ on the real line 
($\tau \in \mathcal{R}$) in such a way that 
$r(\tau) r(\tau') = r(\tau + \tau')$. Set
\begin{displaymath}
\vartheta(\tau,\sigma,p) 
= \frac{\partial \xi(\tau,\sigma,p)}{\partial \sigma}.
\end{displaymath}
From (\ref{9}), (\ref{11}) and (\ref{10}) one gets 
\begin{equation}\label{0e}
\xi(0,0,p) = 0, \, \, \xi(\tau,0,p) = 0,
\end{equation}

\begin{equation}\label{1par}
\xi(\tau, \tau',p) + \xi(\tau + \tau',p) 
= \xi(\tau', \tau'', r(-\tau)p) +
\xi(\tau, \tau' + \tau'',p).
\end{equation}
Now we derive the expression with respect to $\tau''$ at $\tau'' 
= 0$. This yields
\begin{equation}\label{16}
\vartheta(\tau + \tau', 0, p) = \vartheta(\tau', 0,r(-\tau)p) 
+ \vartheta(\tau,\tau',p).
\end{equation}
Let us define now
\begin{displaymath}
\zeta(\tau,p)= \int_{0}^{\tau} \vartheta(\sigma, 0,p) \, {\ud}\sigma =
\int_{0}^{1} \tau \vartheta(\mu\tau,0,p) \, {\ud}\mu.
\end{displaymath}
We have then
\begin{displaymath}
- \Delta[\zeta] = \zeta(\tau + \tau',p) - \zeta(\tau,p) 
- \zeta(\tau',r(-\tau)p)=
\end{displaymath}

\begin{displaymath}
= \int_{0}^{\tau'} \{ \vartheta(\tau + \sigma,0,p) 
- \vartheta(\sigma,0,r(-\tau)p) \} \, {\ud}\sigma.
\end{displaymath}
Using now the Eq. (\ref{16}) and (\ref{0e}) we get
\begin{displaymath}
- \Delta[\zeta] = \int_{0}^{\tau'} 
\vartheta(\tau,\sigma,p) \, {\ud}\sigma 
= \int_{0}^{\tau'} \frac{\partial \xi(\tau,\sigma,p)}{\partial \sigma} \, 
{\ud}\sigma = \xi(\tau, \tau',p)
\end{displaymath}
and $\xi$ is equivalent to 0. 

\vspace{1ex}

Let us recall that the continuous curve $r(\tau)$ in a Lie group $G$ is 
a one-parameter subgroup if and only if $r(\tau_{1})r(\tau_{2}) = 
r(\tau_{1} + \tau_{2})$ \emph{i.e.} $r(\tau) = (r_{0})^{\tau}$, for
some element $r_{0} \in G$, note that the real power $r^{\tau}$ 
is well defined on a Lie group (at least
on some neighborhood of $e$). The coordinates $\rho^{k}$ in $G$ 
are \emph{canonical} if and only if any curve of the form 
$r(\tau) = \tau \rho^{k}$ (where the coordinates $\rho^{k}$ are 
fixed) is a one-parameter subgroup 
(the curve $r(\tau) = \tau \rho^{k}$ 
will be denoted in short by $\tau a$, with the coordinates 
of $a$ equal to $\rho^{k}$). The "vector" $a$ is called by 
physicists the \emph{generator} of the one-parameter
subgroup $\tau a$. 

A local exponent $\xi$ of a Lie group $G$ is called \emph{canonical} if 
$\xi(r,s,p)$ is differentiable 
in all variables and $\xi(r,s,p) = 0$ 
if $r$ and $s$ are elements of the same one-parameter subgroup.

\vspace{1ex}

\begin{lem} Every local exponent $\xi$ of a Lie group is equivalent 
to a canonical local exponent. \end{lem}\label{can}

\vspace{1ex}

\emph{Proof}. Set $\rho^{j}$ and $\sigma^{i}$ for the canonical 
coordinates of the two elements $r,s \in G$ respectively, and
define 
\begin{displaymath}
\vartheta_{k} = \frac{\partial \xi(r,s,p)}{\partial\sigma^{k}}.
\end{displaymath}
Let us define now
\begin{displaymath}
\zeta(r,p) = \int_{0}^{1} \sum_{k=1}^{n} \rho^{k} 
\vartheta_{k}(\mu r,0,p) \, {\ud}\mu.
\end{displaymath}
Consider a one-parameter subgroup $r(\tau)$ generated by $a$, 
\emph{i.e}
$r(\tau) = \tau a$. Because $\xi$ is a local exponent 
fulfilling (\ref{9}), (\ref{10}) and (\ref{11}) then
$\xi_{0}(\tau, \tau',p) \equiv \xi(\tau a, \tau' a,p)$ fulfills 
(\ref{0e}) and (\ref{1par}). Repeating now the same 
steps as in the 
proof of Lemma 2 one can show that 
\begin{displaymath}
\xi(\tau a, \tau' a, p) + \Delta[\zeta(\tau a,p)] = 0. 
\end{displaymath} 

\vspace{1ex}

\begin{lem} Let $\xi$ and $\xi'$ be two differentiable and equivalent 
local exponents of a Lie group $G$, and assume $\xi$ to 
be canonical. Then $\xi'$ is canonical if and only if $\xi' = \xi 
+ \Delta[\Lambda]$, where $\Lambda(r,p)$ is 
a linear form in the 
canonical coordinates of $r$ fulfilling the condition
that
$\Lambda(a,(\tau a)p)$ is constant as a function of $\tau$,
i.e. it follows that\end{lem}\label{eqcan}
\begin{equation}\label{condLem}
\frac{d\Lambda(a, (\tau a)p)}{d\tau} 
= \lim_{\epsilon \to 0}\frac{\Lambda(a,(\epsilon a)p) 
- \Lambda(a,p)}{\epsilon} = 0.
\end{equation}

\vspace{1ex}

The  limit in the above expression can be understood in the ordinary 
point-wise sense with respect to $p$. But after this the assertion of 
the Lemma is much stronger then (\ref{condLem}). We will use later 
the fact that (\ref{condLem}) is true in any linear topology in the
function linear space (with the obvious addition) of $\theta(p)$  
providing that $p \to \Lambda(a, p)$ is differentiable in the sense 
of this linear topology. In the sequel we will use the simple notation 
\begin{displaymath}
\boldsymbol{a} f(p) = \frac{df((\tau a)p)}{d\tau}\Big\vert_{\tau = 0}
= \lim_{\epsilon \to 0}\frac{f((\epsilon a)p) - f(p)}{\epsilon},
\end{displaymath}
and 
$\boldsymbol{a}f(p) = 0$ means that $f(p)$ is constant along 
the integral curves $p(\tau) = (\tau a) p_{0}$. After this
from 
the condition of Lemma 6 follows that 
\begin{displaymath}
\boldsymbol{a}\Lambda(a,p) =0.
\end{displaymath}

\vspace{1ex}

\emph{Proof} of Lemma 4. 1$^{o}$. Necessity of the condition. Because 
the exponents are equivalent we have 
$\xi'(r,s,p) = \xi(r,s,p) +\Delta[\zeta]$. Because both $\xi$ and 
$\xi'$ are differentiable then $\zeta(r,p)$ also is 
a differentiable 
function, which follows from Lemma 1. Suppose that $r=\tau a$ and 
$s= \tau' a$. Because
of both $\xi$ and $\xi'$ are \emph{canonical} 
we have $\xi(\tau a, \tau' a,p) = \xi'(\tau a, \tau' a,p) = 0$, such 
that $\Delta[\zeta](\tau a, \tau' a,p) = 0$, \emph{i.e.}
\begin{displaymath}
\zeta((\tau + \tau')a, p) = \zeta(\tau a, p) 
+ \zeta(\tau' a,(-\tau a)p).
\end{displaymath}  
Applying recurrently this formula one gets
\begin{equation}\label{17}
\zeta(\tau a,p) = \sum_{k=0}^{n-1} \zeta(\frac{\tau}{n}a,
(-\frac{k}{n}\tau a)p).
\end{equation}
$\zeta$ is differentiable (up to any order) and we can use the Taylor 
Theorem. Because in addition $\zeta(0,p) =0$ 
we get the following 
formula 
\begin{displaymath}
\zeta(\frac{\tau}{n}a,p) = \zeta'(0,p)\frac{\tau}{n} + 
\frac{1}{2}\zeta''(\theta_{\frac{\tau}{n}}\frac{\tau}{n}a,p)
\Big(\frac{\tau}{n}\Big)^{2},
\end{displaymath}
where $\zeta'$ and $\zeta''$ stand for the first and the second derivative 
of $\zeta(xa,p)$ with
respect to $x$, and $0 \leq \theta_{\frac{\tau}{n}} 
\leq 1$. Recall that
in the Taylor formula
$f(x+h) = f(x) + f'(x)h + 1/2f''(x + \theta_{h} h) h^{2}$ the 
$\theta_{h} \in [0,1]$ depends on $h$, which is marked by 
the subscript $h$: $\theta_{h}$. Inserting $\tau =n =1$ to the formula and 
multiplying it by $\tau/n$ (provided
the coordinates $a$ of an element 
$r_{0} \in G$ are chosen in such a way that $r_{0}$
lies in the neighborhood ${\mathfrak{N}}_{0}$ on which the exponents 
$\xi$ and $\xi'$ are defined)
one gets
\begin{displaymath}
\frac{\tau}{n}\zeta(a,p) = \frac{\tau}{n}\{\zeta'(0,p) 
+ \frac{1}{2} \zeta''(\theta_{1}a,p)\}.
\end{displaymath} 
Taking now the difference of the last two formulas we get
\begin{displaymath}
\zeta(\frac{\tau}{n}a,p) = \frac{\tau}{n}\Big\{ \zeta(a,p) 
- \frac{1}{2} \zeta''(\theta_{1}a,p) \Big\} + 
\frac{1}{2}\Big(\frac{\tau}{n} \Big)^{2} 
\zeta''(\theta_{\frac{\tau}{n}} \frac{\tau}{n}a,p).
\end{displaymath}
Inserting this to the formula (\ref{17}) we get
\begin{displaymath}
\zeta(\tau a,p) = \frac{\tau}{n} \sum_{k=0}^{n-1} 
\Big\{\zeta(a,(-\frac{k}{n}\tau a)p) - 
\frac{1}{2}\zeta''(\theta_{1}a, (-\frac{k}{n}\tau a)p) \Big\}  
\end{displaymath}

\begin{displaymath}
+ \frac{1}{2}\Big(\frac{\tau}{n}\Big)^{2} \sum_{k=0}^{n-1} 
\zeta''(\theta_{\frac{\tau}{n}}\frac{\tau}{n}a, (-\frac{k}{n}\tau a)p).
\end{displaymath}
Denote the supremum and the infimum of the function $\zeta''(xa, (-ya)p)$ 
in the square
$(0 \leq x \leq \tau, 0 \leq y \leq \tau)$ by $M$ and $N$ 
respectively. We have
\begin{displaymath}
\frac{1}{2}\Big(\frac{\tau}{n}\Big)^{2}nN 
+ \frac{\tau}{n}\sum_{k=0}^{n-1} \Big\{\zeta(a,(-\frac{k}{n}a)p) 
- \frac{1}{2}\zeta''(\theta_{1} a,(-\frac{k}{n}\tau a)p) \Big\} 
\end{displaymath}

\begin{displaymath}
\leq \zeta(\tau a,p) \leq \frac{1}{2}\Big(\frac{\tau}{n}\Big)^{2}nM + 
\end{displaymath}

\begin{displaymath}
+ \frac{\tau}{n}\sum_{k=0}^{n-1} \Big\{\zeta(a,(-\frac{k}{n}\tau a)p) 
- \frac{1}{2}\zeta''(\theta_{1}a,(-\frac{k}{n}\tau a)p) \Big\}.
\end{displaymath}
Passing to the limit $n \rightarrow + \infty$ we get
\begin{displaymath}
\zeta(\tau a,p) = \int_{0}^{\tau} \Big\{\zeta(a,(-\sigma a)p) 
- \frac{1}{2}\zeta''(\theta_{1}a, (-\sigma a)p) \Big\} \, {\ud}\sigma.
\end{displaymath}
Taking into account that the functions $\zeta'$ and $\zeta''$ are 
independent the general solution $\zeta$
fulfilling $\Delta[\zeta]
(\tau a, \tau' a, p) = 0$ for any $\tau$, $\tau'$, $a$ and any 
$p\in \mathcal{M}$, can be written in the following form
\begin{equation}\label{18}
\zeta(\tau a,p) = \int_{0}^{\tau} \varsigma(a,(-\sigma a)p) \, {\ud} \sigma,
\end{equation}
where $\varsigma = \varsigma(r,p)$ is any differentiable function. 
Differentiate now the expression (\ref{18}) with respect to $\tau$ at 
$\tau = 0$. After this one gets
\begin{equation}\label{19}
\varsigma(a, p) = \sum_{k=1}^{n}\lambda_{k}(p)a^{k}, \, \, \, 
\textrm{with} \, \, \, \lambda_{k}(p) = 
\frac{\partial \zeta(0,p)}{\partial a^{k}},
\end{equation}
where $a^{k}$ stands for the coordinates of $a$. So, the function 
$\varsigma(a,p)$ is linear with respect to $a$. 
Suppose that the 
spacetime
coordinates $X$ are chosen in such a way that the integral 
curves $p(x) = (xa)p_{0}$ are coordinate lines, 
which is possible for appropriately small $\tau$. There are of course 
three remaining families of coordinate lines
beside $p(x)$, which can be 
chosen in arbitrary way, the parameters of which will be denoted by $y_{i}$.
After this, Eq. (\ref{18}) reads
\begin{displaymath}
\zeta(\tau a,x,y_{i}) = \int_{0}^{\tau} \varsigma(a,x 
- \sigma, y_{i}) \, {\ud}\sigma.
\end{displaymath} 
So, because $\varsigma(a,p)$ is linear with respect to $a$, then for 
appropriately small $a$ one gets
\begin{displaymath} 
\zeta(a,x,y_{i}) = \frac{1}{\tau} \int_{0}^{\tau} \varsigma(a,x 
- \sigma, y_{i}) \, {\ud} \sigma =
\end{displaymath}

\begin{displaymath}
= \frac{1}{\tau} \int_{x - \tau}^{x} \varsigma(a,z,y_{i}) \, {\ud}z,
\end{displaymath} 
for any $\tau$ (of course with appropriately small $\vert \tau \vert$, 
in our case $\vert \tau \vert \leq 1$) and for any 
(appropriately small) 
$x$.
But this is possible for the function $\varsigma(a,x,y_{k})$ 
continuous in $x$  (in our case differentiable in $x$) 
if and only if $\varsigma(a,x,y_{k})$ does not depend on $x$. This means 
that $\zeta(a,x,y_{k})$ 
does not depend on $x$ and the condition of 
Lemma 6 is proved.
\\ $2^{o}$. Sufficiency of the condition is trivial.

\section{The Lie Algebras}

According to Lemma 3 we can assume that the 
exponent is canonical. Also we confine ourselves to the subgroup of 
$\{\theta(p),r\} \in  H$ with differentiable $\theta$, and denote 
this subgroup by the same letter $H$. We embed this subgroup $H$ in an 
infinite dimensional Lie group with the manifold structure modeled on 
a Banach space. We will extensively use the theory developed by Birkhoff 
\cite{Birkhoff} and Dynkin \cite{Dynkin}. For the systematic treatment of 
manifolds modeled on Banach spaces, see e.g. \cite{Lang}.  
By this embedding we ascribe bi-uniquely a Lie algebra to the group $H$ with
the convergent Baker-Hausdoff series.  

Note first that the formula
\begin{displaymath}
H \times \mathcal{L}^{2}(\mathcal{M},\mu,\mathcal{H})\ni 
(\{\theta(p),r \}, \phi) \to e^{i\theta(p)}T_{r}\phi
\end{displaymath}
together with (\ref{H-action}) can be viewed as rule
giving the action of $H$ in the direct integral Hilbert space
$\int_{\mathcal{M}} \mathcal{H}_{p} \, d\mu(p)$ defined in the 
${\bf 3}^{rd}$ section. Moreover, this is a unitary action, provided
$\mu$ is $G$-invariant. In accordance to \cite{Birkhoff} the group 
$D$ of \emph{all} unitary operators of a Hilbert space is an infinite 
dimensional Lie group. By this $H = N \circledS G$ can be viewed as a 
subgroup of a Lie group. 

We consider now the closure $\overline{H}$ of $H$ in the sense of the 
topology in $D$. 

\vspace{1ex}

\begin{lem}The subgroup $\overline{H}$ also has locally the 
structure of the semi-direct product $\overline{N} \circledS G$.\end{lem}

\vspace{1ex}

\emph{Proof}. It is the consequence of the following four facts.

(1)      $\overline{N}$ is a normal subgroup of $\overline{H}=
         \overline{N\centerdot G}$.

(2)      $G$ is finite dimensional, so, $\overline{G} = G$.	 
	   
(3)      Locally (in a neighborhood $\mathcal{O}$) the multiplication 
         in $D$ is given by the Baker-Hasdorff formula in the Banach algebra of $D$. 
         Because $\overline{N}$ is  normal in $\overline{H}$, then the above mapping 
         converts locally  the multiplication $\overline{N}\centerdot S$ of 
         $\overline{N}$ by any subset $S$ of $\overline{H}$ into the sum $\overline{N} 
         + S$. Because $G$ is finite dimensional, and by this is locally
         compact, the  neighborhood $\mathcal{O}$ can bee chosen in such a 
         way that locally (in the closure of $\mathcal{O} + \mathcal{O}$)
         \begin{displaymath}
          \overline{N}+\overline{G} 
          = \overline{N+G} = \overline{H}.
         \end{displaymath}	  

(4)      The local $\overline{N}$ (intersected with $\overline{\mathcal{O}}$) 
	 has finite co-dimension in local $\overline{N+G}$ (intersected with
	 $\overline{\mathcal{O} + \mathcal{O}}$) and by this it 
	 splits locally $\overline{N+G}$. So, we have locally, \emph{i.e.} 
	 in $\overline{\mathcal{O} + \mathcal{O}}$
	 \begin{displaymath}
	  \overline{N} + \overline{G}
	  = \overline{H} = \overline{N}\oplus G', 
	 \end{displaymath}
	  where $\overline{G'} = G'$ and  $\oplus$ stands for direct sum.
	  From this it follows that $G' =\overline{G}$ locally.

\vspace{2ex}

Because $\overline{H}= \overline{N}\circledS G$ every $h\in \overline{H}$
is uniquely representable in the form $ng$, where $n \in \overline{N}$ and 
$g \in G$. Notice now that
\begin{displaymath}
(n_{1}g_{1})(n_{2}g_{2}) = n_{1}g_{1}n_{2}g_{1}^{-1}g_{1}g_{2} =
[n_{1}(g_{1}n_{2}g_{1}^{-1})](g_{1}g_{2})
\end{displaymath}
and that $g_{1}n_{2}g_{1}^{-1} \in \overline{N}$ because $\overline{N}$
is normal in $\overline{H}$. Let us denote the automorphism 
$n \to ghg^{-1}$ of $\overline{N}$ by $R_{g}$. 
The group $\overline{H}$ can be locally viewed as a topological product of 
Banach spaces $\overline{\mathfrak{N}}\times 
\mathfrak{G}$  one of which, namely $\mathfrak{G}$ is
finite dimensional and isomorphic tho the Lie algebra of $G$.
The multiplication in $\overline{H}$ can be written as
$(n_{1}, g_{1})(n_{2},g_{2}) = (n_{1}R_{g_{1}}(n_{2}), g_{1}g_{2})$. 
Moreover, $\overline{N}$ can be viewed locally as the Banach space 
$\overline{\mathfrak{N}}$ with the multiplication law given by the vector 
addition in $\overline{\mathfrak{N}}$.   

Now, our task is to reconstruct the Lie algebra $\overline{\mathfrak{H}}$
corresponding to the subgroup $\overline{H}$. 
  
Let $\lambda \to \lambda a$ be a one-parameter subgroup of $G$.
The mapping 
\begin{displaymath}
(\lambda, n) \to (R_{\lambda a}n, \lambda a)
\end{displaymath}
of the Banach space $\mathcal{R} \times \overline{\mathfrak{N}}$ 
into the Banach space $\overline{\mathfrak{N}} \times \mathfrak{G}$ is 
continuous. Indeed. The
multiplication law in $\overline{H}$ is continuous.
On the other hand multiplying $(0,\lambda a)$ and $(n, 0)$
we get $(R_{\lambda a}n, \lambda a)$ from which the continuity
of the above mapping follows. 
By this $\mathcal{R}\ni \lambda \to R_{\lambda a}n \in 
\overline{\mathfrak{N}}$ as well as $\overline{\mathfrak{N}} 
\ni n \to R_{\lambda a}n$ are continuous. By this the function 
$\lambda \to R_{\lambda a}n$
can be integrated over any compact interval and we have 
\begin{displaymath}
R_{\lambda a}\int_{0}^{\tau} R_{\sigma a}n \, d\sigma
= \int_{0}^{\tau}R_{\lambda a} \circ R_{\sigma a}n \, d\sigma
= \int_{0}^{\tau}R_{(\lambda a)(\sigma a)}n \, d\sigma
= \int_{0}^{\tau}R_{(\lambda + \sigma) a}n \, d\sigma.
\end{displaymath}
If one takes it into account, then a straightforward computation
shows that
\begin{displaymath}
    \tau \to (n_{\tau a}, \tau a) 
    := \Big(\int_{0}^{\tau}R_{\sigma a}n \, \ud \sigma, \tau a\Big) 
\end{displaymath}
is a one-parameter subgroup of $\overline{H}$. It is not hard to show
that to any element $h$  of $\overline{H}$ we can construct in 
this way a subgroup passing trough $h$. 
Moreover, the element $\check{a}$  of the algebra $\overline{\mathfrak{H}}$ 
corresponding to the one-parameter subgroup is equal (compare 
\cite{Birkhoff}, \cite{Dynkin})
\begin{equation}\label{vecal}
\lim_{\tau \to 0} \frac{(n_{\tau a}, \tau a)}{\tau}
=\lim_{\tau \to 0} \frac{(\int_{0}^{\tau}R_{\sigma a}n \, 
\ud \sigma, \tau a)}{\tau} = (n, a),
\end{equation}
because for any Banach-valued continuous function 
$\mathcal{R} \ni \sigma \to F(\sigma)$, 
\begin{displaymath}
\lim_{\tau \to 0} \frac{1}{\tau} \int_{0}^{\tau}F(\sigma) \, \ud \sigma
= F(0)
\end{displaymath} 
in the sense of limit induced by the norm in the Banach space. 

The Lie bracket $[\check{a}, \check{b}]$ in  $\overline{\mathfrak{H}}$ 
is uniquely determined by the one-parameter subgroups 
$\tau \check{a} = \{\alpha_{\tau a}, \tau a\}$ and  $\tau\check{b}
= \{\beta_{\tau b}, \tau b\}$ corresponding to $\check{a}$ and $\check{b}$ 
respectively (compare \cite{Birkhoff}, \cite{Dynkin})
\begin{displaymath}
[\check{a},\check{b}] = \lim_{\tau \to 0}
\frac{(\tau\check{a})(\tau\check{b})(\tau
\check{a})^{-1}
(\tau\check{b})^{-1}}{\tau^{2}},
\end{displaymath} 
where the limit is in the sense of the topology induced from the Lie group 
$D$, of course. 

The elements of $H \subset \overline{H}$ are representable in the 
ordinary form $\{\alpha,r\}$ with 
differentiable $\alpha = \alpha(p), p\in \mathcal{M}$, and $r \in G$. 
Let $\lambda \to \lambda a$ be a one-parameter subgroup of $G$.
Consider the above defined operator $R_{\lambda a}$. Its restriction
to $H \subset \overline{H}$ is given by (remember that $\xi$ is canonical) 
\begin{displaymath}
\alpha(p) \to (R_{\lambda a}\alpha)(p) = \alpha((\lambda a)^{-1}p).
\end{displaymath}
We compute now explicitly the Lie bracket and the Jacobi identity
for all the elements $(\alpha(p), a)$ of the subalgebra $\mathfrak{H} \subset 
\overline{\mathfrak{H}}$ corresponding to the subgroup $H$.    
A rather straightforward computations, in which the continuity 
of $(\sigma a,\alpha) \to R_{\sigma a}\alpha$ as well as the homomorphism 
property $R_{\tau a}R_{\lambda b} = R_{(\tau a)(\lambda b)}$ are 
used, gives 
\begin{equation}\label{20'}
[\check{a}, \check{b}] = \{\boldsymbol{a}\beta - \boldsymbol{b}\alpha 
+ \Xi(a,b, p), [a,b] \},
\end{equation}
  
\begin{displaymath}
\Xi(a,b,p) = \lim_{\tau \to 0} \tau^{-2}\{\xi((\tau a)(\tau b), 
(\tau a)^{-1}(\tau b)^{-1},p) + 
\end{displaymath}

\begin{equation}\label{20''}
+ \xi(\tau a, \tau b, p) + \xi((\tau a)^{-1},(\tau b)^{-1}, 
(\tau b)^{-1}(\tau a)^{-1}p) \},
\end{equation}
Let us stress once more that  
\begin{equation}\label{strong}
\boldsymbol{a}\theta(p) = \lim_{\epsilon \to 0}\frac{\theta((\epsilon a)p) 
- \theta(p)}{\epsilon},
\end{equation}
and the limit is in the sense of topology induced from the Lie group $D$.
   
From the associative law in $H$ one gets
\begin{displaymath}
((\tau \check{a})(\tau \check{b}))(\tau \check{c}) 
= (\tau \check{a})((\tau \check{b})(\tau \check{c})).
\end{displaymath}
We divide now the above expression by $\tau^{3}$ and then pass to the limit 
$\tau \to 0$. Inserting the explicit values we get
\begin{displaymath}
\Xi([a,a'],a'',p) + \Xi([a',a''],a,p) + \Xi([a'',a],a',p) =
\end{displaymath} 

\vspace{-0.5cm}

\begin{equation}\label{21}
= \boldsymbol{a}\Xi(a',a'',p) + \boldsymbol{a'}\Xi(a'',a,p) 
+ \boldsymbol{a''}\Xi(a,a',p),
\end{equation}
which can be shown to be equivalent to the Jacobi identity
\begin{equation}\label{21'}
[[\check{a}, \check{a}'], \check{a}''] + [[\check{a}',\check{a}''],\check{a}] 
+
 [[\check{a}'',\check{a}], \check{a}'] = 0. 
\end{equation}
So, we have reconstructed in this way the Lie algebra 
$\overline{\mathfrak{H}}$ giving explicitly $[\check{a}, \check{b}]$ for all 
$\check{a},\check{b} \in \mathfrak{H} \subseteq \overline{\mathfrak{H}}$. 
Because $\mathfrak{H}$ is dense in $\overline{\mathfrak{H}}$, the local 
exponent $\Xi$ determines the algebra $\overline{\mathfrak{H}}$ uniquely. 
But from the theory of Lie groups the correspondence between the algebras 
$\overline{\mathfrak{H}}$ and local Lie groups $\overline{H}$ is bi-unique, 
at last locally, see e.g. \cite{Birkhoff} and \cite{Dynkin}. So we get  

\vspace{1ex}

{\bf Corollary 1} \, \, \emph{The correspondence $\overline{H} \to 
\overline{\mathfrak{H}}$ between the local group $\overline{H}$ and 
the algebra 
$\overline{\mathfrak{H}}$ is one-to-one}.

\vspace{1ex}

Our method is most effective in the case in which the limit in (\ref{strong}) can
be replaced by the ordinary point-wise limit (with respect to the variable 
$p \in \mathcal{M}$). Then the operator $\boldsymbol{a}$ becomes to be an 
ordinary differential operator. In other words, this is the case when the 
existence of the limit in $\overline{\mathfrak{H}}$ implies the existence of 
the point-wise limit and the both limits
are always equal.We describe the important case of this situation. Suppose, that the
subalgebra $\mathfrak{H}'$ generated from the set of elements $\{0,a\}$, where 
$a$ is any element of algebra of $G$, is $\emph{finite dimensional}$. Then the 
topology induced in $\mathfrak{H}'$ from $\overline{\mathfrak{H}}$ is equivalent 
to any Hausdorff linear topology in $\mathfrak{H}'$. In particular it can be the 
point-wise topology.

The natural question arises, then, when the group $G$ possess finite 
dimensional extended algebra $\mathfrak{H}'$. We show now that this 
is always the case in the nonrelativistic case.       

\vspace{1ex}

In the nonrelativistic theory $\xi=\xi(r,s,t)$ depends on the time. 
In this case, according to our assumption about $G$, any $r \in G$ 
transforms simultaneity hyperplanes into simultaneity hyperplanes. So, 
there are two
possibilities for any $r \in G$. First, when $r$ does not 
change the time: $t(rp) = t(p)$ and the second in
which the time is 
changed, but in such a way that $t(rp) - t(p) = f(t)$. \emph{We assume 
in addition that the base generators $a_{k} \in \mathfrak{G}$ can be 
chosen in such a way that only one acts on the time as the translation
and the remaining ones do not act on the time}. 
Because we are searching a finite dimensional extension
we can assume that the operators $\boldsymbol{a}$ are the ordinary 
differential operators. After this the Jacobi identity (\ref{21}) reads
\begin{equation}\label{Jac1}
\Xi([a,a'],a'') + \Xi([a', a''],a) + \Xi([a'',a],a') 
= \partial_{t} \Xi(a',a''),
\end{equation}
if one and only one among $a,a',a''$ is the time translation generator, 
namely $a$, and 
\begin{eqnarray}\label{Jac2}
\Xi([a,a'],a'') + \Xi([a',a''], a) + \Xi([a'',a],a') = 0,
\end{eqnarray} 
in all remaining cases. The Jacobi identity (\ref{Jac1}) and (\ref{Jac2}) 
can be treated as a system of ordinary differential linear equations for 
the finite set of unknown functions $\Xi_{ij}(t) = \Xi(a_{i}, a_{j},t)$, 
where $a_{i}$ is the base in the Lie algebra of $G$. Indeed, the identity 
gives us the only set of nontrivial equations
which provides us the tool 
for the classification of possible $\Xi$-s on $G$, or equivalently 
the possible algebras $\mathfrak{H}$.
Some of the unknowns $\Xi$ are not 
determined by the Jacobi equations (in general), and some $\Xi_{ij}(t)$ 
are left
completely arbitrary. In section \ref{examples} we will show  
that different values of undetermined $\Xi_{ij}$ lead to homomorphic
algebras. Then, we can put
the undetermined $\Xi_{ij}$ equal to zero, 
and do not lost any generality. After this we are left with
a system of 
fewer equations for a fewer set of unknowns $\Xi$, which has to be determined. 
Let
us order the fewer set of unknowns $\Xi$ and compose a vector-column 
$\boldsymbol{y}$ of unknowns.
For a fixed $t$ any $\boldsymbol{y}$ is an 
element of a finite dimensional vector space $Y$.
Then, the system of linear equations can be written as follows   
\begin{equation}\label{Jacexp}
\dot{\boldsymbol{y}} = \boldsymbol{Ay},
\end{equation}
where dot is the time derivation and $\boldsymbol{A}$ is a linear operator 
in $Y$.
From this system of linear equations  we see that the time derivative 
$\partial_{t}\Xi_{ij}$ is determined by linear combinations of $\Xi_{ij}$. 
From this follows that $\Xi_{ij}$ compose
 the base for the algebra $\mathfrak{H}$, which shows the finite 
dimensionality of $\mathfrak{H}$. This simplifies the classification theory 
for time dependent $\xi$,
when the the only transformation acting on the time 
is the time translation. 
However, the reasoning fails in general and it is an open question
if a given group $G$ possess a finite dimensional extension ascribed to the
exponent $\Xi$ in question.

\section{Classification of Local Exponents of Lie Groups}

Because the exponent $\xi$ determines the multiplication rule in $H$ and 
vice-versa, then from the Corollary 1 of the preceding section it follows

\vspace{1ex}

{\bf Corollary 2} \, \, \emph{The correspondence $\xi \to \Xi$ between the 
local $\xi$ and the infinitesimal exponent $\Xi$ is one-to-one}.

\vspace{1ex}

Note that the words 'local $\xi = \xi(r,s,p)$' mean that $\xi(r,s,p)$ is 
defined for $r$ and $s$ belonging
 to a fixed neighborhood 
${\mathfrak{N}}_{0} \subset G$ of $e \in G$, but 
\emph{in our case it is defined 
globally as a function of the 
spacetime variable $p \in \mathcal{M}$}.

\vspace{1ex}

Consider the nonrelativistic theory for the moment.
Suppose the dimension of $G$ to be $n$. Let $a_{k}$ with $k \leq n$ be the 
base in the Lie algebra
 $\mathfrak{G}$ of $G$. Let us introduce the base 
$\check{a}_{j}$ in $\mathfrak{H}$
 in the following way: 
$\check{a}_{n+1} = \{\alpha_{1}(t), 0\}, \ldots , \check{a}_{n+q} = 
\{\alpha_{q}(t), 0\}$ and $\check{a}_{1} = \{0, a_{1}\},
\ldots , \check{a}_{n} = \{0, a_{n}\}$.
After this we have
\begin{equation}\label{raycom}
[\check{a}_{i},\check{a}_{j}] = c_{ij}^{k}\check{a}_{k} + \Xi(a_{i},a_{j}),
\end{equation}
for $i,j \leq n$. It means that,
in general, the commutation relations of a \emph{ray} representation of 
$G$ are not equal to
the commutation relations $[A_{i}, A_{j}] 
= c_{ij}^{k}A_{k}$ of $G$, but they are equal to
$[A_{i},A_{j}] = c_{ij}^{k}A_{k} + \Xi(a_{i}, a_{j},t) \centerdot {\bf 1}$. 
The generator $A_{i}$ corresponding to $a_{i}$ is defined in the following 
way\footnote{The transformation $T_{r}$ does not act in the ordinary Hilbert 
space but in the Hilbert bundle space $\mathcal{R}\triangle\mathcal{H}$,
by this we cannot immediately appeal to the Stone and 
G{\aa}rding Theorems. Nonetheless, $T_{r}$ induces a unique unitary 
representation acting in the Hilbert space 
$\int_{\mathcal{R}} \mathcal{H}_{t} d\mu(t)$ 
and it can be shown that it is meaningful to tell 
about the generators $A$ of $T_{r}$.} 
\begin{displaymath}
A_{i}\psi = \lim_{\tau \to 0} \frac{(T_{\tau a_{i}} 
- {\boldsymbol{1}})\psi}{\tau}.
\end{displaymath}

\vspace{1ex}

Now, we pass to describe the relation between the infinitesimal exponents 
$\Xi$ and 
local exponents $\xi$. Let us compute first the infinitesimal 
exponents $\Xi$ and $\Xi'$ given by (\ref{20''})
which correspond to the two equivalent canonical local exponents $\xi$ and 
$\xi' = \xi + \Delta[\Lambda]$.
 Inserting $\xi' = \xi + \Delta[\Lambda]$ 
to the formula (\ref{20''}) one gets
\begin{equation}\label{24}
\Xi'(a,b,p) = \Xi(a,b,p) +  \boldsymbol{a}\Lambda(b,p) 
- \boldsymbol{b}\Lambda(a,p) - \Lambda([a,b],p). 
\end{equation}     
Recall, that according to the Lemma 3, we can confine ourselves to the 
canonical exponents.
 According to Lemma 4 $\Lambda = \Lambda(a, (\tau b)p)$ 
is a constant function of $\tau$ if 
$a = b$, and $\Lambda(a,p)$ is linear 
with respect to $a$ (we use the canonical coordinates on $G$). 
By this $\Xi'(a,b,p)$ is antisymmetric in $a$ and $b$ and fulfills 
(\ref{21}) if only $\Xi(a,b,p)$
 is antisymmetric in $a$ and $b$ and fulfills 
(\ref{21}). 
This suggests the definition: \emph{two infinitesimal exponents 
$\Xi$ and $\Xi'$ will be called
 equivalent if and only if the relation} 
(\ref{24}) \emph{holds}. For short we write the relation (\ref{24}) 
as follows: 
\begin{displaymath}
\Xi' = \Xi + d[\Lambda].
\end{displaymath}

\vspace{1ex}

\begin{lem}Two canonical local exponents $\xi$ and $\xi'$ are equivalent 
if and only
 if the corresponding infinitesimal exponents $\Xi$ and $\Xi'$ 
are equivalent. \end{lem}

\vspace{1ex}

\emph{Proof}. (1) Assume $\xi$ and $\xi'$ to be equivalent. Then, by the 
definition of equivalence of infinitesimal
exponents $\Xi' = \Xi + d[\Lambda]$. (2) Assume $\Xi$ and $\Xi'$ to be 
equivalent: $\Xi' = \Xi + d[\Lambda]$ for some linear form $\Lambda(a,t)$ 
such that $\Lambda(a,(\tau a)p)$ does not depend on $\tau$. 
Then $\xi + \Delta[\Lambda] \to \Xi'$, and by the uniqueness of the 
correspondence $\xi \to \Xi$ (Corollary 2),
 $\xi' = \xi + \Delta[\Lambda]$, 
\emph{i.e.} $\xi$ and $\xi'$ are equivalent. 

\vspace{1ex}

At last from Lemma 3 every local exponent is equivalent to a canonical 
one and 
by the Corollary 2 to every $\Xi$ corresponds uniquely a local 
exponent. So, we can
 summarize our results in the following

\vspace{1ex}

\begin{twr} (1) On a Lie group G, every local exponent $\xi(r,s,p)$ 
is equivalent to a canonical
 local exponent $\xi'(r,s,p)$  which, 
on some canonical neighborhood ${\mathfrak{N}}_{0}$, is analytic
in canonical coordinates 
of r and s and and vanishes if r and s belong 
to the same one-parameter subgroup. Two canonical
local exponents $\xi,\xi'$ are equivalent if and only if 
$\xi' = \xi + \Delta[\Lambda]$ on some canonical
neighborhood, where $\Lambda(r,p)$ is a linear form in the canonical 
coordinates of $r$ such that
 $\Lambda(r,sp)$ does not depend on $s$
if $s$ belongs to the same one-parameter subgroup as $r$.
(2) To every canonical
 local exponent of $G$ corresponds uniquely an 
infinitesimal exponent $\Xi(a,b,p)$ on the Lie
 algebra $\mathfrak{G}$ of $G$, 
i.e. a bilinear antisymmetric form which satisfies  the identity 
$\Xi([a,a'],a'',p) +\Xi([a',a''],a,p)+ \Xi(a'',a],a',p) 
= \boldsymbol{a}\Xi(a',a'',p) + \boldsymbol{a'}\Xi(a'',a,p) +
\boldsymbol{a''}\Xi(a,a',p)$. The correspondence is linear. (3) Two canonical 
local exponents 
$\xi,\xi'$ are equivalent if and only if the corresponding 
$\Xi,\Xi'$ are equivalent, i.e. 
$\Xi'(a,b,p) = \Xi(a,b,p) 
+ \boldsymbol{a}\Lambda(b,p) - \boldsymbol{b}\Lambda(a,p) - \Lambda([a,b],p)$
where $\Lambda(a,p)$ is a linear form in $a$ on $\mathfrak{G}$ such that 
$\tau \to \Lambda(a,(\tau b)p)$ is constant if $a = b$.
(4) There exist a one-to-one correspondence between the equivalence classes 
of local exponents
 $\xi$ (global in $p \in \mathcal{M}$) of $G$ and the 
equivalence classes of infinitesimal exponents $\Xi$ of 
$\mathfrak{G}$. \end{twr}

\section{Global Extensions of Local Exponents}

Theorem 3 provides the full classification of exponents $\xi(r,s,p)$ local 
in $r$ and $s$, defined for all $p \in \mathcal{M}$.
But we will show that if $G$ is connected and simply connected,  
then we have the following theorems.
 (1) If an extension $\xi'$ of a given 
local (in $r$ and $s$) exponent $\xi$ does exist, then it is uniquely 
determined (up to 
the equivalence transformation (\ref{13})) (Theorem 4). 
(2) There exists such an extension $\xi'$ (Theorem 5),
proved for $G$ which possess finite dimensional extension
$\mathfrak{H}'$ only.

In the global analysis the topology of $H$ induced from $D$ is not
applicable. For we are not able to prove that the homomorphism
(\ref{izo}) is continuous when $\xi$ is not canonical. Note, that 
any $\xi$ is equivalent to a canonical one, but only \emph{locally}!
We introduce another topology. Because of the semidirect structure
of $H = N\circledS G$ it is sufficient to introduce it into
$N$ and $G$ separately in such a manner that $G$ acts continuously 
on $N$, compare e.g. \cite{Mackey}. From the discussion of section
\ref{exponents} it is sufficient to introduce the Fr\'echet
topology of almost uniform convergence in the function space $N$. 
Indeed, from the strong continuity of $\xi$ and $\zeta$ in
(\ref{izo}) it follows that the multiplication rule as well as
the homomorphism (\ref{izo}) are continuous. 

\vspace{1ex}

\begin{twr} Let $\xi$ and $\xi'$ be two equivalent local exponents of a 
connected and simply 
connected group $G$, so that $\xi' = \xi 
+ \Delta[\zeta]$ on some neighborhood, and assume the exponents 
$\xi_{1}$ and $\xi_{1}'$ of $G$ to be extensions of $\xi$ and $\xi'$ 
respectively. Then, for all $r,s \in G$
$\xi_{1}'(r,s,p) = \xi_{1}(r,s,p) + \Delta[\zeta_{1}]$ where $\zeta_{1}(r,p)$ 
is strongly
 continuous in $r$ and differentiable in $p$, and $\zeta_{1}(r,p) 
= \zeta(r,p)$, for all $p \in \mathcal{M}$ and for all 
$r$ belonging to some neighborhood of $e \in G$. \end{twr}
 
\vspace{1ex}  

\emph{Proof}. The two exponents $\xi_{1}$ and $\xi_{1}'$ being 
\emph{strongly continuous} (by assumption) 
define two semicentral extensions $H_{1} = N_{1} \circledS G$ and 
$H_{1}'= N'_{1}\circledS G$, which are continuous groups. 
Note, that the linear groups $N_{1},N'_{1}$ are connected and 
simply connected.
 Because $H_{1}$ and $H_{1}'$ both
are semi-direct products of two connected
and simply connected groups they are both connected and simply connected. 
Eq. (\ref{izo}) defines a local isomorphism mapping $h: \check{r} \to 
\check{r}' = h(\check{r})$ 
of $H_{1}$ into $H_{1}'$
\begin{displaymath}
h(\check{r}) = h(\theta,r) = \{\theta(p) - \zeta(r,p),r\}
\end{displaymath}
on the appropriately small neighborhood of $e$ in $G$, on which 
$\xi_{1} = \xi$
 and $\xi_{1}' = \xi'$. Because $H_{1}$ and $H_{1}'$ are 
connected and simply connected
 the isomorphism $h$ given by (\ref{izo}) can 
be uniquely extended to an isomorphism 
$h_{1}(\check{r}) = h(\theta,r) 
= \check{r}'$ of the entire groups $H_{1}$ and $H_{1}'$ such 
that $h_{1}(\check{r}) = h(\check{r})$ on some neighborhood of $H_{1}$, see
\cite{Pontrjagin}, Theorem 80. The isomorphism $h_{1}$ defines
an isomorphism of the two abelian subgroups  $N_{1}$ and $h_{1}(N_{1})$. 
By (\ref{izo})
 $h_{1}(\theta,e) = \{\theta, e\}$ locally in $H_{1}$, that is 
for $\theta$ lying appropriately 
close to 0 (in the metric sense defined 
previously). Both $N_{1}$ and $h_{1}(N_{1})$ are connected, and $N_{1}$ is 
in addition simply connected,
so applying once again the Theorem 80 of \cite{Pontrjagin}, one can see that
$h_{1}(\theta,e) = \{\theta,e\}$ for all $\theta$. Set $h_{1}(0,r) 
= \{ - \zeta_{1}(p),g(r)\}$.
Note, that because $f_{1}$ is an isomorphism it is continuous in 
the topology of $H_{1}$ and $H_{1}'$. By this $\zeta_{1}(r,p)$ is 
\emph{strongly
 continuous} in $r$ and $g(r)$ is a continuous function of $r$. 
The equation $\{\theta,r\} = \{\theta,e\}\{0,r\}$ implies that 
$h_{1}(\theta(p),r) = \{\theta(p) - \zeta_{1}(r,p), g(r)\}$.
Computing now $h_{1}(0,r)h_{1}(0,s)$ we find that $g(rs) = g(r)g(s)$. So, 
$g(r)$ is an automorphism of a connected and simply connected $G$, for which
$g(r) = r$ locally, then applying once more the Theorem 80 of 
\cite{Pontrjagin} one shows that
 $g(r) =r$ for all $r$. Thus
\begin{displaymath}
h_{1}(\check{r}) = h_{1}(\theta(p),r) = \{\theta(p) - \zeta_{1}(r,p),r\},
\end{displaymath}   
for all $\check{r} \in H_{1}$. Finally, $h_{1}(0,r)h_{1}(0,s) 
= h_{1}(\xi_{1}(r,s,p),rs)$. Hence
\begin{displaymath}
 \{\xi_{1}'(r,s,p) - \zeta_{1}(r,p) - \zeta_{1}(s,r^{-1}p), rs\} =
\end{displaymath}

\begin{displaymath}
 \{\xi_{1}(r,s,p) - \zeta_{1}(rs,p), rs\},
\end{displaymath}
for all $r,s,p$. That is, $\xi_{1}'(r,s,p) = \xi_{1}(r,s,p) 
+ \Delta[\zeta_{1}]$ for all $r,s,p$ and 
by (\ref{izo}) $\zeta_{1}(r,p) =\zeta(r,p)$ on some neighborhood of 
$e$ on $G$.

\vspace{1ex}

The following Theorem is proved for the group $G$ having
 a finite dimensional
extended algebra $\mathfrak{H}'$.

\vspace{1ex}

\begin{twr} Let $G$ be connected and simply connected Lie group. Then to 
every exponent $\xi(r,s,X)$ of $G$ defined locally in $(r,s)$ there exists an exponent 
$\xi_{0}$ of $G$ defined on the whole group $G$ which is an extension of $\xi$. If $\xi$
is differentiable, $\xi_{0}$ may be chosen differentiable. \end{twr}

\vspace{1ex}

Because the proof of Theorem 5 is identical as that of the Theorem 5.1 in 
\cite{Bar}, 
we do not present it explicitly\footnote{In the proof we consider 
the finite dimensional extension $H'$ of $G$ instead of the Lie group
$H$ in the proof 
presented in \cite{Bar}. The remaining replacements 
are rather trivial, but we 
mark them here explicitly to simplify the reading. 
(1) instead of the formula 
$\bar{r}' = \bar{t}(\theta)\bar{r} 
= \bar{r}\bar{t}(\theta)$ of  (5.3) in \cite{Bar}
we have $\check{r}' 
= \check{t}(\theta(r^{-1}p))\check{r} = \check{r}\check{t}(\theta(p))$. 
By this, from the formula 
$(\check{h}_{1}(r)\check{h}_{1}(s))\check{h}_{1}(g)$ 
$= \check{h}_{1}(r)(\check{h}_{1}(s)\check{h}_{1}(g))$
(see \cite{Bar}) follows 
$\xi(r,s,p) + \xi(rs,g,p)$ $= \xi(s,g,r^{-1}p) + \xi(r,sg,p)$ instead of (5.8) 
in \cite{Bar}.
(2) Instead of (4.9), (4.10) and (4.11) we use the Iwasawa-type 
construction presented in this paper.
(3) Instead of Lemma 4.2 in \cite{Bar} 
we use the Lemma 1. 
}. Note that the 
proof largely
 rests on the global theory of classical (finite dimensional) 
Lie groups. Namely,
 it rests on the theorem that there always exists the 
universal covering group to
 any finite dimensional Lie group. We can use 
those methods because there exist a finite dimensional extension $H'$
of 
$G$.

We have obtained the full classification of
 time dependent $\xi$ 
defined on the whole group $G$ for Lie groups $G$ which are connected 
and simply connected in the nonrelativistic
theory.  But for any Lie group $G$ there exists
the universal covering group $G^{*}$ which is connected and simply connected. 
So, for $G^{*}$
the correspondence $\xi \to \Xi$ is one-to-one, that is, 
to every $\xi$ there exists the unique $\Xi$
 and vice versa, to every $\Xi$ 
corresponds uniquely $\xi$ defined on the whole group $G^{*}$ 
and the 
correspondence preserves the equivalence relation. Because $G$ and $G^{*}$ 
are locally
isomorphic the infinitesimal exponents $\Xi$'s are exactly the 
same for $G$ and for $G^{*}$.
 Because to every $\Xi$ there does exist 
exactly one $\xi$ on $G^{*}$, so, if there
 exists the corresponding $\xi$ 
on the whole $G$ to a given $\Xi$, then such a $\xi$ is unique.
We have obtained in this way the full classification of $\xi$ defined
on a whole Lie group $G$ for any Lie group $G$, in the sense that no $\xi$
 can be omitted in the classification. The set of equivalence classes of 
$\xi$ is considerably smaller than the set of equivalence classes of $\Xi$, 
it may happen that to some local $\xi$ there does not exist any global 
extension. Therefore, the classification is full in this sense in the 
relativistic theory also.

Take, for example, a Lie subgroup $G$ of the Milne group and its ray 
representation $T_{r}$. 
We have classified in this way all exponents for 
this
 $T_{r}$ and $r \in G$. In general such a $\Xi$ may exists that there 
does not exist any 
$\xi$ corresponding to this $\Xi$ if the group $G$ is 
not connected and simply connected.
 But this not important for us, 
the important fact is that no $\xi(r,X)$ can be omitted in this 
classification.

\section{Examples}\label{examples}

\subsection{Example 1: The Galilean Group}

According to the conclusions of section \ref{motivation} one should 
\emph{a priori} investigate such representations of 
the Galilean group $G$ 
which fulfill the Eq. (\ref{rowt}), with $\xi$ depending on the time. 
The following paradox, then, arises. Why the transformation law $T_{r}$ under 
the Galilean group has time-independent $\xi$ in (\ref{rowt}) independently of 
the fact if it is a covariance group or a symmetry group? We will solve the 
paradox in this subsection. Namely, we will show that any representation
of the Galilean group fulfilling (\ref{rowt}) is equivalent to a 
representation
 fulfilling (\ref{rowt}) with time-independent $\xi$. 
This is a rather peculiar property of the Galilean group not valid in 
general. For example, this is not true for
the group of Milne transformations.

According to section \ref{generalization} we shall determine all equivalence 
classes of infinitesimal exponents $\Xi$ of the Lie algebra $\mathfrak{G}$ of 
$G$
to classify all $\xi$ of $G$. The commutation relations for the Galilean 
group are as follows
\begin{equation}\label{26a}
[a_{ij},a_{kl}] = \delta_{jk}a_{il} - \delta_{ik}a_{jl}+\delta_{il}a_{jk}-
\delta_{jl}a_{ik}, 
\end{equation}

\begin{equation}\label{26b}
[a_{ij},b_{k}] = \delta_{jk}b_{i} - \delta_{ik}b_{j}, \, [b_{i},b_{j}] = 0,
\end{equation}

\begin{equation}\label{26c}
[a_{ij.d_{k}}] = \delta_{jk}d_{i} - \delta_{ik}d_{j}, \, [d_{i},d_{j}] 
= 0, [b_{i},d_{j}] = 0,
\end{equation}

\begin{equation}\label{26d}
[a_{ij},\tau] = 0, [b_{k},\tau] = 0, [d_{k},\tau] = b_{k},
\end{equation}
where $b_{i},d_{i}$ and $\tau$ stand for the generators of space translations, 
the proper Galilean transformations and time translation respectively and 
$a_{ij} = - a_{ji}$ are rotation generators. Note, that the Jacobi identity 
(\ref{Jac2}) is identical to the Jacobi identity in the ordinary
Bargmann's Theory of time-independent exponents (see \cite{Bar}, Eqs (4.24) 
and (4.24a)). So, using (\ref{26a}) -- (\ref{26c}) we can proceed exactly 
after Bargmann (see \cite{Bar}, pages 39,40)
and show that any infinitesimal 
exponent 
defined on the subgroup generated by $b_{i}, d_{i}, a_{ij}$ is 
equivalent to an exponent
equal to zero with the possible exception of 
$\Xi(b_{i},d_{k},t) = \gamma \delta_{ik}$, where 
$\gamma = \gamma(t)$. So, 
the only components of $\Xi$ defined on the whole algebra $\mathfrak{G}$ which 
can a priori be not 
equal to zero are: $\Xi(b_{i},d_{k},t) = \gamma \delta_{ik}, \, \Xi(a_{ij}, 
\tau,t), \,
\Xi(b_{i}, \tau,t)$ and $\Xi(d_{k},\tau,t)$. We compute first 
the function $\gamma(t)$. Substituting
$a= \tau, \, a' = b_{i}, a'' = d_{k}$ 
to (\ref{Jac1}) we get ${\ud}\gamma/{\ud}t = 0$, so that $\gamma$
is a constant, we denote the constant value of $\gamma$ by $m$. Inserting 
$a = \tau, \, a' = a_{i}^{s},
\, a'' = a_{sj}$ to (\ref{Jac1}) and summing 
up with respect to $s$ we get $\Xi(a_{ij},\tau,t) = 0$. 
In the same way, but with the substitution $a = \tau, a' = a_{i}^{s}, a'' 
= b_{s}$, one shows that
$\Xi(b_{i},\tau, t) = 0$. At last the substitution 
$a=\tau, a' = a_{i}^{s}, a''=d_{s}$ to (\ref{Jac1})
and summation with respect to $s$ gives $\Xi(d_{i},\tau,t) = 0$. 
We have proved in this way that any time depending $\Xi$ on $\mathfrak{G}$ 
is equivalent to a time-independent one. In other words, we get a 
one-parameter family of possible $\Xi$, 
with the parameter equal to the 
inertial mass $m$ of the system in question. 
Any infinitesimal time-dependent 
exponent  of the Galilean group is equivalent
to the above time-independent 
exponent $\Xi$ with some value of the parameter $m$; and any two infinitesimal
exponents with different values of $m$ are inequivalent.   
As was argued in \ref{generalization} (Theorems 3 $\div$ 5) the 
classification of $\Xi$ gives the full classification of $\xi$.
Moreover, it can be shown that the classification of $\xi$ 
is equivalent to the classification of possible $\theta$-s 
in the transformation law
\begin{equation}\label{trapsi}
T_{r}\psi(p) = e^{i\theta(r,p)}\psi(r^{-1}p)
\end{equation}
for the spinless nonrelativistic particle.
On the other hand, the exponent $\xi(r,s,t)$ of the representation $T_{r}$ 
given by (\ref{trapsi}) 
can be easily computed to be equal 
$\theta(rs,p - \theta(r,p - \theta(s,r^{-1}p$,
and the infinitesimal exponent belonging to $\theta$ defined as 
$\theta(r,p = -m\vec{v}\centerdot \vec{x} + \frac{m}{2}\vec{v}^{2}t$,
covers the whole one-parameter family of the classification 
(its infinitesimal exponent is equal to that 
infinitesimal exponent $\Xi$,
which has been found above). So, the standard $\theta(r,p = - m\vec{v} 
\centerdot \vec{x} + \frac{m}{2}\vec{v}^{2}t$, 
covers the full 
classification of possible $\theta$-s in (\ref{trapsi}) for the Galilean 
group. Inserting the standard
form for $\theta$ we see that $\xi$ does not 
depend on $X$ but only on $r$ and $s$. By this, any time-depending
$\xi$ on $G$ is equivalent to a time-independent one. 

This result can be obtained in the other way. Namely, using now the Eq. 
(\ref{raycom})
we get the commutation relations for the \emph{ray} 
representation $T_{r}$ of the Galilean group 
\begin{displaymath}
[A_{ij},A_{kl}] = \delta_{jk}A_{il} - \delta_{ik}A_{jl} - \delta_{jl}A_{ik},  
\end{displaymath}

\begin{displaymath}
[A_{ij},B_{k}] = \delta_{jk}B_{i} - \delta_{ik}B_{j}, \, [B_{i},B_{j}] = 0,
\end{displaymath}

\begin{displaymath}
[A_{ij}, D_{k}] = \delta_{jk}D_{i} - \delta_{ik}D_{j}, 
\end{displaymath}

\begin{displaymath}
[D_{i},D_{j}] =0, \, [B_{i},D_{j}] = m\delta_{ij},
\end{displaymath}

\begin{displaymath}
[A_{ij}, T] = 0, \, [B_{k},T] = 0, \, [D_{k}, T] = B_{k},
\end{displaymath} 
where the generators $A_{ij}, \ldots $ which correspond to the generators 
$a_{ij}, \ldots$ of the one-parameter subgroups $r(\sigma) = \sigma a_{ij}, 
\ldots$ are defined in the following way (compare the $4^{th}$ footnote)
\begin{displaymath}
A_{ij}\psi(X) = \lim_{\sigma \to 0} \frac{(T_{r(\sigma)} 
- \boldsymbol{1})\psi(X)}{\sigma}.
\end{displaymath}
$A_{ij}$ is well defined for any differentiable $\psi(p)$. So, we get the 
standard commutation relations such as in the case when the Galilean group 
is a symmetry group.
The above standard commutation relations for the 
transformation
$T_{r}$ of the form (\ref{trapsi}) gives a differential 
equations for $\theta$. It it easy to show, that they can be solved
uniquely (up to an irrelevant function $f(t)$ of time and the group 
parameters) and the solution has the standard form
$\theta(r,p)= - m\vec{v}\centerdot \vec{x} + f(t)$.

Note, that to any $\xi$ (or $\Xi$) there exists a corresponding $\theta$ 
(and  such a $\theta$ is unique up to a trivial equivalence relation). 
As we will see this is not the case for the 
Milne group, where such 
$\Xi$ do exist which cannot be realized by any $\theta$.

\subsection{Example 2: Milne group as a covariance group}

In this subsection we apply the theory of section \ref{generalization} to 
the Milne transformations group. We proceed 
like with the Galilean group 
in the preceding section. The Milne group $G$ does not form
any Lie group, which complicates the situation. We will go on according 
to the following plan. First, 1) we define
the topology in the Milne group. 
Second, 2) we define the sequence $G(1) \subset \ldots \subset G(m) \subset 
\ldots$ of Lie subgroups of the Milne group $G$ dense in $G$. 3) Then we 
compute the infinitesimal exponents
and exponents for each $G(m)$, 
$m = 1,2, \ldots $, and by this the $\theta$ in (\ref{trapsi}) for $G(m)$.
4) As we have proved in \ref{generalization} the (strong) continuity of 
the exponent $\xi(r,s,t)$ in the group variables
follows as a consequence 
of the Theorem 2.
It can be shown that also $\theta(r,p)$ is strongly 
continuous in the group variables
$r \in G$  By this, $\theta(r,p)$ defined 
for $r \in G(m), m = 1, 2, \ldots $
can be uniquely extended on the whole 
group $G$. This can be done effectively thanks to
the assumption that the 
wave equation is local.

Before we go further on we make an important remark. The Milne group $G$ 
is an infinite dimensional group and 
there are infinitely many ways in 
which a topology can be introduced in $G$. On the other hand the physical 
contents
of the continuity assumption of section \ref{generalization} 
depends effectively on the topology in $G$. By this the assumption is in 
some sense empty.
True, but it is important to stress here, that the whole 
relevant physical content
rests on the Lie subgroup $G(m)$ 
(see the further text for the definition of $G(m)$) for a sufficiently 
large $m$, and not on the whole $G$. That is,
the covariance condition with 
respect
to $G(m)$ for sufficiently large $l$, instead of $G$ is sufficient 
for us. By this, there are no ambiguities in the continuity assumption. 
The topology in $G$ is not important from the physical 
point of view, and the extension of the formula (\ref{trapsi}) from $G(m)$ 
to the whole group is of secondary importance. 
However, we construct such an extension to make our considerations more 
complete, living the opinion
about the "naturality" of this extension 
to the reader.    

1) Up to now the Milne group of transformations  
\begin{equation}\label{tra}
(\vec{x},t) \to (R\vec{x} + \vec{A}(t), t + b),
\end{equation}
where $R$ is an orthogonal matrix, and $b$ is constant,
has not been strictly defined. The extent of arbitrariness of the function 
$\vec{A}(t)$ in
(\ref{tra}) has been left open up to now. The topology depends 
on the degree of this arbitrariness. 
It is natural to assume the function 
$\vec{A}(t)$ in (\ref{tra}) to be differentiable up to any order. Consider
the subgroups $G_{1}$ and $G_{2}$ of the Milne group which consist of the 
transformations:
$(\vec{x},t) \to (\vec{x}+\vec{A}(t),t)$ and $(\vec{x},t) 
\to (R\vec{x}, t + b)$ respectively. 
Then the Milne group $G$ is equal to 
the semidirect product $G_{1}\centerdot G_{2}$,
where $G_{1}$ is the normal 
factor of $G$. It is sufficient to introduce a topology in $G_{1}$ and then 
define
the topology in $G$ as the semi-Cartesian product topology, where it 
is clear what is the topology in the Lie group
$G_{2}$. We introduce a linear 
topology in the linear group $G_{1}$ which makes it a Fr\'echet space,
in which the time derivation operator $\frac{{\ud}}{{\ud}t}: \vec{A} 
\to \frac{{\ud}\vec{A}}{{\ud}t}$ becomes  a continuous operator. Let $K_{N}, 
N = 1, 2, \ldots$ be such a sequence of compact sets of $\mathcal{R}$, that
\begin{displaymath}
K_{1} \subset K_{2} \subset \ldots \, \, \, \textrm{and} \, \, \, \bigcup_{N} 
K_{N} = {\mathcal{R}}.
\end{displaymath} 
Then we define a separable family of seminorms
\begin{displaymath} 
p_{N}(\vec{A}) = \max  \big\{ \big\vert \vec{A}^{(n)}(t) \big\vert, 
t \in K_{N}, n \leq N \big\},    
\end{displaymath}
where $\vec{A}^{(n)}$ denotes the  $n$-th order time derivative of $\vec{A}$. 
Those seminorms define 
on $G_{1}$ a locally convex metrizable topology. 
For example, the metric
\begin{displaymath}
d(\vec{A_{1}},\vec{A_{2}}) = \max_{N \in {\mathcal{N}}} 
\frac{2^{-N}p_{N}(\vec{A_{2}} - \vec{A_{1}})}{1 + p_{N}(\vec{A_{2}} 
- \vec{A_{1}})}
\end{displaymath}
defines the topology. 

2) It is convenient to rewrite the Milne transformations (\ref{tra}) in the 
following form
\begin{displaymath}
\vec{x'} = R\vec{x} + A(t) \vec{v}, \, \, \, t' = t + b,
\end{displaymath}
where $\vec{v}$ is a constant vector, which does not depend on the time $t$. 
We define
the subgroup $G(m)$ of $G$ as the group of the following 
transformations
\begin{displaymath}
\vec{x'} = R\vec{x} + \vec{v}_{(0)} + t\vec{v}_{(1)} 
+ \frac{t^{2}}{2!}\vec{v}_{(2)} + \ldots + \frac{t^{m}}{m!}\vec{v}_{(m)},
\, \, \, t' = t + b, 
\end{displaymath}
where $R = (R_{a}^{b}), v_{(n)}^{k}$ are the group parameters -- 
in particular the group $G(m)$ has the dimension
equal to $3m + 7$. 

3) Now we investigate the group $G(m)$, that is, we classify their 
infinitesimal exponents. The
commutation relations of $G(m)$ are 
as follows
\begin{equation}\label{Milne1}
[a_{ij},a_{kl}] = \delta_{jk}a_{il} - \delta_{ik}a_{jl} 
+ \delta_{il}a_{jk} - \delta_{il}a_{ik},
\end{equation}

\begin{equation}\label{Milne2}
[a_{ij},d_{k}^{(n)}] = \delta_{jk}d_{i}^{(n)} 
- \delta_{ik}d_{j}^{(n)}, \, [d_{i}^{(n)},d_{j}^{(k)}] = 0, 
\end{equation}

\begin{equation}\label{Milne3}
 [a_{ij},\tau] = 0, \,  [d_{i}^{(0)},\tau]=0, \, [d_{i}^{(n)},\tau] 
= d_{i}^{(n-1)}, 
\end{equation}
where $d_{i}^{(n)}$ is the generator of the transformation 
$r(v_{(n)}^{i})$:
\begin{displaymath}
{x'}^{i} = x^{i} + \frac{t^{n}}{n!}v_{(n)}^{i},
\end{displaymath}
which will be called the $n$-acceleration, especially 0-acceleration is 
the ordinary space translation. All the relations (\ref{Milne1}) and 
(\ref{Milne2}) are identical with (\ref{26a}) $\div$ (\ref{26c}) with
the $n$-acceleration instead of the Galilean transformation. So, the 
same argumentation as that used for the Galilean group gives: 
$\Xi(a_{ij}, a_{kl})= 0$, $\Xi(a_{ij},d_{k}^{(n)}) = 0$, and 
$\Xi(d_{i}^{(n)}, d_{j}^{(n)}) = 0$. Substituting $a_{i}^{h}, a_{hi}, 
\tau$ for $a,a',a''$ into the Eq. (\ref{Jac1}),
making use of the 
commutation relations and summing up with respect to $h$ we get 
$\Xi(a_{ij},\tau) = 0$. Substituting $a_{i}^{h}, d_{h}^{(l)}, 
d_{k}^{(n)}$ for $a,a',a''$ into the Eq. (\ref{Jac2})
we get in the analogous way $\Xi(d_{i}^{(l)}, d_{k}^{(n)}) 
= \frac{1}{3}\Xi(d^{(l)h},d_{h}^{(n)}) \, \delta_{ik}$.
Substituting $a_{i}^{h}, d_{h}^{(n)}, \tau$ for $a,a',a''$ into 
the Eq. (\ref{Jac1}), making use of commutation relations,
and summing up with respect to $h$, we get $\Xi(d_{i}^{(n)}, \tau) = 0$. 
Now, we substitute 
$d_{k}^{(n)}, d_{i}^{(0)}, \tau$ for $a,a',a''$ in 
(\ref{Jac1}), and proceed recurrently with respect to $n$,
we obtain in this way $\Xi(d_{i}^{(0)},d_{k}^{(n)}) 
= P^{(0,n)}(t)\delta_{ik}$, where $P^{(0,n)}(t)$ is a polynomial
of degree $n-1$ -- the time derivation of $P^{(0,n)}(t)$ has to be 
equal to $P^{(0,n-1)}(t)$, and $P^{(0,0)}(t) = 0$. 
Substituting $d_{k}^{(n)}, d_{i}^{(l)}, \tau$ to (\ref{Jac1}) 
we get in the same way 
$\Xi(d_{k}^{(l)}, d_{i}^{(n)}) = P^{(l,n)}(t)\delta_{ki}$, where 
$\frac{{\ud}}{{\ud}t}P^{(l,n)} = P^{(l-1,n)} + P^{(l,n-1)}$.
This allows us to determine all $P^{(l,n)}$ by the recurrent integration 
process. Note that $P^{(0,0)} = 0$, and 
$P^{(l,n)} = - P^{(n,l)}$, so we 
can compute all $P^{(1,n)}$ having given the $P^{(0,n)}$. Indeed, we have 
$P^{(1,0)} = - P^{(0,1)}, P^{(1,1)} = 0, {\ud}P^{(1,2)}/{\ud}t = P^{(0,2)} 
+ P^{(1,1)}, {\ud}P^{(1,3)}/{\ud}t = P^{(0,3)} + P^{(1,2)},
\ldots$ and after $m-1$ integrations we compute all $P^{(1,n)}$. Each 
elementary integration introduces a new
independent parameter (the arbitrary 
additive integration constant). Exactly in the same way we can compute
all $P^{(2,n)}$ having given all $P^{(1,n)}$ after the $m-2$ elementary 
integration processes. In general
the $P^{(l-1,n)}$ allows us to compute 
all $P^{(l,n)}$ after the $m-l$ integrations. So, $P^{(l,n)}(t)$ are 
$l+n - 1$-degree polynomial functions of $t$, and all are determined by 
$m(m+1)/2$ integration constants. Because 
$d[\Lambda](d_{i}^{(n)}, d_{k}^{(l)}) = 0$, the exponents $\Xi$ defined 
by different polynomials $P^{(l,n)}$
are inequivalent. By this the space 
of inequivalent classes of $\Xi$ is $m(m+1)/2$-dimensional.

However, not all $\Xi$ can be realized by the transformation $T_{r}$ of 
the form (\ref{trapsi}). All
the above integration constants have to be equal 
to zero with the exception of those in $P^{(0,n)}(t)$. 
By this, all exponents of $G(m)$, which can be realized by the 
transformations $T_{r}$ of the form
(\ref{trapsi}) are determined by the 
polynomial $P^{(0,m)}$, that is, by $m$ constants. We show it first
for the group $G(2)$ , because the case is the simplest one and it 
suffices to explain
the principle of all computations for all $G(m)$. 
From the above analysis 
we have $P^{(0,1)} = \gamma_{1}, P^{(0,2)} = 
\gamma_{1}t + \gamma_{2},
P^{(1,2)} = \frac{1}{2}\gamma_{1}t^{2} 
+ \gamma_{2}t + \gamma_{(1,2)}$, where $\gamma_{i}, \gamma_{(1,2)}$
are the integration constants. We will show that $\gamma_{(1,2)} = 0$. 
A simple computation gives the following
formula $\xi(r,s) = \theta(rs,X) 
- \theta(r,X) - \theta(s,r^{-1}X)$ for the exponent of the representation 
$T_{r}$ of the form (\ref{trapsi}). Inserting this $\xi$ to the Eq. 
(\ref{20''})
and performing a rather straightforward computation we get 
the following formula 
\begin{displaymath}
 \Xi(d_{i}^{(k)}, d_{j}^{(n)}) = \frac{t^{n}}{n!}
\frac{\partial^{2}\theta}{\partial x^{j}\partial v_{(k)}^{i}} - 
\frac{t^{k}}{k!}\frac{\partial^{2}\theta}{\partial x^{i}\partial v_{(n)}^{j}}, 
\end{displaymath}
for the infinitesimal exponent $\Xi$ of the representation $T_{r}$ given 
by (\ref{trapsi}),
where the derivation with respect to $v_{(p)}^{q}$ is 
taken at $v_{(p)}^{q} = 0$. Comparing this $\Xi(d_{i}^{(k)}, d_{j}^{(n)})$ 
with $P^{(k,n)}\delta_{ij}$ we get the equations
\begin{equation}\label{thetaXi}
\frac{t^{n}}{n!}\frac{\partial^{2}\theta}{\partial x^{j}\partial v_{(k)}^{i}} 
- 
\frac{t^{k}}{k!}\frac{\partial^{2}\theta}{\partial x^{i}\partial v_{(n)}^{j}} 
= P^{(k,n)} \delta_{ij}.
\end{equation}
Because of the linearity of the problem,
we can consider the three cases $1^{o}$. $\gamma_{(2)} = \gamma_{(1,2)} = 0$, 
$2^{o}$. $\gamma_{1} = \gamma_{(1,2)} = 0$ and $3^{o}$. $\gamma_{1} 
= \gamma_{2} = 0$, separately.
In the case $1^{o}$. we have the solution 
\begin{displaymath} 
\theta(r,X) = \gamma_{1}\frac{{\ud}\vec{A}}{{\ud}t}\centerdot \vec{x} 
+ \widetilde{\theta}(t),
\end{displaymath}
where $\widetilde{\theta}(t)$ is an arbitrary function of time and the group 
parameters, and $\vec{A}(t) \in G(2)$. 
In the case $2^{o}$ we have
\begin{displaymath}
\theta(r,X) = \gamma_{2}\frac{{\ud}^{2}\vec{A}}{{\ud}t^{2}} \centerdot \vec{x} 
+ \widetilde{\theta}(t), 
\end{displaymath}
with arbitrary function $\widetilde{\theta}(t)$ of time.
Consider at last 
the case $3^{o}$. From (\ref{thetaXi}) we have (corresponding to 
$(k,n) = (0,1), (0,2))$ 
and $(1,2)$ respectively) 
\begin{equation}\label{theta1}
t\frac{\partial^{2}\theta}{\partial x^{j}\partial v_{(0)}^{i}} 
- \frac{\partial^{2}\theta}{\partial x^{i}\partial v_{(1)}^{j}} = 0, 
\end{equation}

\begin{equation}\label{theta2}
\frac{t^{2}}{2}\frac{\partial^{2}\theta}{\partial x^{j}\partial v_{(0)}^{i}} 
- 
\frac{\partial^{2}\theta}{\partial x^{i}\partial v_{(2)}^{j}} = 0,
\end{equation}

\begin{equation}\label{theta3}
\frac{t^{2}}{2}\frac{\partial^{2}\theta}{\partial x^{j}\partial v_{(1)}^{i}} 
-
t \frac{\partial^{2}\theta}{\partial x^{i}\partial v_{(2)}^{j}} 
= \gamma_{(1,2)}\delta_{ij}.
\end{equation}
From (\ref{theta3}) and (\ref{theta2}) we get
\begin{equation}\label{theta4}
\frac{t^{2}}{2} \big\{\frac{\partial^{2}\theta}{\partial x^{j}
\partial v_{(1)}^{i}}
 - t \frac{\partial^{2}\theta}{\partial x^{j}\partial v_{(0)}^{i}} \big\} 
= \gamma_{(1,2)} \delta_{ij}.
\end{equation}
But $\Xi(d_{i}^{(0)}, d_{j}^{(0)}) = 0 = \partial^{2}\theta/\partial x^{j}
\partial v_{(0)}^{i}
- \partial^{2}\theta/\partial x^{i}\partial v_{(0)}^{j}$, 
so, from (\ref{theta4}) and (\ref{theta1}) we get
\begin{displaymath}
0 = \frac{\partial^{2}\theta}{\partial x^{i}\partial v_{(0)}^{j}}
\big\{\frac{t^{3}}{2} - \frac{t^{3}}{2}\big\}
= \gamma_{(1,2)} \delta_{ij},
\end{displaymath}
and $\gamma_{(1,2)} = 0$.

The following   
\begin{equation}\label{thetaG(2)}
\theta(r,X) = \gamma_{1}\frac{{\ud}\vec{A}}{{\ud}t} \centerdot \vec{x} + 
\gamma_{2}\frac{{\ud}^{2}\vec{A}}{{\ud}t^{2}} \centerdot \vec{x} 
+ \widetilde{\theta}(t)
\end{equation}
fulfills all Eqs. (\ref{thetaXi}) with $k,n \leq 2$ and its local exponents 
cover the full classification
of $\Xi$'s for $G(2)$ which can be realized by 
$T_{r}$ of the form (\ref{trapsi}), that is, all $\Xi$'s with
$\gamma_{(1,2)} = 0$. Then, the formula (\ref{thetaG(2)}) gives the most 
general 
$\theta$ in (\ref{trapsi}) for $r \in G(2)$.  This is because the 
classification of $\Xi$'s covers
the classification of all possible $\theta$'s 
(however we live it without proof). 

It can be immediately seen that any integration constant $\gamma_{(l,q)}$ of 
the polynomial
$P^{(l,q)}(t)$ has to be equal to zero if $l,q \neq 0$, 
provided the exponent $\Xi$ belongs to the
representation $T_{r}$ of the form 
(\ref{trapsi}). The argument is essentially the same
as that for $\gamma_{(1,2)}$. 
It is sufficient to consider (\ref{thetaXi}) 
for the four cases: $(k,n) = (l-1,q-1), (l-1,q), (l,q)$
and $(l,q-1)$ respectively. Because of the linearity of the considered 
problem, it is sufficient to consider
the situation with the integration constants 
in $P^{(k,n)}$ equal to zero with the possible exception of the 
integration constant $\gamma_{(l,q)}$. We get in this way the equations 
(\ref{thetaXi}) corresponding to $(l-1,q-1),(l-1,q),
(l,q)$ and $(l, q-1)$ with the right hand sides equal to zero with the 
exception of the right hand side of the 
equations corresponding to 
$(k,n) = (l,q)$, which is equal to $\gamma_{(l,q)} \delta_{ij}$. 
From the equations (\ref{thetaXi}) corresponding to $(k,n) = (l,q)$ 
and $(l-1,q)$ we get   
\begin{displaymath}
\frac{t^{q}}{q!}\frac{\partial^{2}\theta}{\partial x^{j}
\partial v_{(l-1)}^{i}} - 
\frac{t^{q+1}}{q!l}\frac{\partial^{2}\theta}{\partial x^{j}
\partial v_{(l-1)}^{i}} = \gamma_{(l,q)}\delta_{ij}.
\end{displaymath}
From this and the equations (\ref{thetaXi}) corresponding to 
$(k,n) = (q-1,l-1)$ we get
\begin{displaymath}
\frac{t^{q}}{q!}\frac{\partial^{2}\theta}{\partial x^{j}
\partial v_{(l)}^{i}} -
\frac{t^{l+1}}{l!q}\frac{\partial^{2}\theta}{\partial x^{i} 
\partial v_{(q-1)}^{j}} = \gamma_{(l,q)}\delta_{ij}.
\end{displaymath}
From this and the equations (\ref{thetaXi}) corresponding to 
$(k,n) = (l,q-1)$ one gets
\begin{displaymath}
0 = \frac{t^{l+1}}{l!q}\frac{\partial^{2}\theta}{\partial x^{i}
\partial v_{(q-1)}^{j}} -
\frac{t^{l+1}}{l!q}\frac{\partial^{2}\theta}{\partial x^{i}
\partial v_{(q-1)}^{j}} = \gamma_{(l,q)} \delta_{ij},
\end{displaymath}
which gives the result that $\gamma_{(l,q)} = 0$. 

Consider the $\theta$, given by the formula
\begin{equation}\label{thetaG(m)}
\theta(r,p) = \gamma_{1}\frac{{\ud}\vec{A}}{{\ud}t} 
+ \gamma_{2}\frac{{\ud}^{2}\vec{A}}{{\ud}t^{2}} + \ldots + 
\gamma_{m}\frac{{\ud}^{m}\vec{A}}{{\ud}t^{m}} + \widetilde{\theta}(t),
\end{equation}
for $r \in G(m)$, where $\gamma_{i}$ are the integration constants which 
define the polynomial 
$P^{(0,m)} = \gamma_{1}\frac{t^{m-1}}{(m-1)!} 
+ \gamma_{2}\frac{t^{(m-2)}}{(m-2)!} + \ldots + \gamma_{m}$, and
$\widetilde{\theta}(t)$ is any function of the time $t$ and eventually 
of the group parameters .
A rather simple computation shows that this 
$\theta$ fulfills all (\ref{thetaXi}) for $k,n \leq m$ and that it covers 
all possible $\Xi$ which can be realized by (\ref{trapsi}). That is, 
the infinitesimal exponents
corresponding to the $\theta$ given by 
(\ref{thetaG(m)}) give all possible $\Xi$ with 
all integration constants $\gamma_{(k,n)} = 0$, for $k,n \neq 0$. 
So, the most general $\theta(r,p)$ defined for $r \in G(m)$ is given by 
(\ref{thetaG(m)}). 

At this place we make use of the assumption that
the wave equation is local. 
It can be shown that (we live it without proof) from this assumption
that the $\theta(r,p)$ can be a function of a finite order derivatives of 
$\vec{A}(t)$, say $k$-th at most,
the higher derivatives cannot enter into 
$\theta$. By this, the most general $\theta(r,p)$ defined
for $r \in G(m)$ 
has the following form 
\begin{equation}\label{thetaG}    
\theta(r,X) = \gamma_{1}\frac{{\ud}\vec{A}}{{\ud}t} + \ldots 
+ \gamma_{k}\frac{{\ud}^{k}\vec{A}}{{\ud}t^{k}} 
+ \widetilde{\theta}(t),
\end{equation}

4) Now, we extend the formula (\ref{thetaG}) on the whole Milne group $G$. 
It is a known fact that the time derivative operator ${\ud}/{\ud}t: \vec{A} 
\to {\ud}\vec{A}/{\ud}t$ is a continuous operator on $G$
in the topology introduced in 1), see e.g. \cite{Rudin}. It remains to 
show that the sequence $G(m), m \in {\mathcal{N}}$
is dense in $G$. The proof 
of this presents no difficulties\footnote{It is sufficient 
to use the following two facts. 1) The Weierstrass Theorem:
\emph{For any continuous (and by this any differentiable) 
function $f(t)$ 
and any compact set $\mathcal{C}$ there exist a sequence of polynomial 
functions 
$P_{n}(t)$, uniformly convergent  to $f(t)$ on $\mathcal{C}$}. 
2) The following Theorem:
\emph{ Let $\{P_{n}(t)\}$ be a sequence of functions 
differentiable in the interval $[a,b]$, convergent
at least in one point of 
this interval. If the sequence $\{P_{n}'(t)\}$ of derived functions is 
uniformly convergent
in $[a,b]$ to the function $\varphi(t)$, then the sequence 
of primitive functions $\{P_{n}(t)\}$ (anti-derivatives of $P_{n}'$)
is uniformly convergent to a differentiable function $\phi(t)$ the derivative 
$\phi'(t)$ of which is equal to $\varphi(t)$
in $[a,b]$}.}.
By this the function 
$\theta(r,p)$ can be uniquely extended on the whole group $r \in G$
\begin{displaymath}
\theta(r,X) = \gamma_{1}\frac{{\ud}\vec{A}}{{\ud}t} + \ldots 
+ \gamma_{4}\frac{{\ud}^{k}\vec{A}}{{\ud}t^{k}} 
+ \widetilde{\theta}(t). 
\end{displaymath}
It should be stressed here that not only the topology in $G$ is needed to 
derive the formula, but
also the locality assumption is very important. If 
the coefficients $a, b^{i}, \ldots , g$ in the wave
equation were admitted 
to be nonlocal, then an infinite number of other solutions for $\theta$ in 
$G$ 
would exist.   

\vspace{1cm}

{\bf ACKNOWLEDGMENTS}

\vspace{0.5cm}

The author is indebted for helpful discussions to A. Staruszkiewicz and A. 
Herdegen. the author is also indebted  to the Referee, who suggest him  
the explicit use of the Hilbert bundle formalism.

The paper was financially supported by the KBN grant no. 
5 P03B 09320.
 Without this help the paper would have never been written.

\end{document}